\documentclass[a4paper, 12pt]{article}
\usepackage[colorlinks = true,
            linkcolor = blue,
            urlcolor  = blue,
            citecolor = blue,
            anchorcolor = blue]{hyperref}
\usepackage[utf8]{inputenc}
\usepackage[english]{babel}
\usepackage[margin=1.2in]{geometry}
\usepackage{amsthm,amsmath,amsfonts,amssymb}
\usepackage[authoryear]{natbib}
\setcitestyle{round}
\usepackage{graphicx}%% uncomment this for including figures
\usepackage{authblk}
\usepackage{orcidlink}
\newcommand{\keywords}[1]{\textit{\textbf{Keywords:} #1}}

\usepackage{subcaption}
\usepackage{graphicx}
\usepackage{stmaryrd} % For double brackets
\usepackage{multicol} % For appendix verbatim

\graphicspath{{figures/}}

\newcommand{\mywidth}{0.8\textwidth}

\newcommand{\traj}{Y}
\newcommand{\trajset}{\mathbf{\traj}}
\newcommand{\trajobs}{y}
\newcommand{\trajsetobs}{\mathbf{\trajobs}}
\newcommand{\latent}{Z}
\newcommand{\latentset}{\mathbf{\latent}}
\newcommand{\latentobs}{z}
\newcommand{\latentsetobs}{\mathbf{\latentobs}}
\newcommand{\covar}{\textsc{x}}
\newcommand{\covarset}{\textbf{\covar}}

\newcommand{\decod}{g}
\newcommand{\decodpar}{\theta}
\newcommand{\varpar}{\phi}
\newcommand{\varmean}{\mu}
\newcommand{\varsd}{\gamma}
\newcommand{\varvar}{\varsd^2}
\newcommand{\encod}{h}

\newcommand{\BC}{\text{BC}}

\newcommand{\DKL}{D_{\text{KL}}}
\newcommand{\VAE}{\text{VAE}}
\newcommand{\CVAE}{\text{CVAE}}
\newcommand{\ELBO}{\text{ELBO}}
\newcommand{\PC}{\text{SI}}
\newcommand{\PClag}{\text{D}\PC}
\newcommand{\PCall}{\text{G}\PC}
\newcommand{\barPClag}[1]{\overline{\PClag}_{#1}}
\newcommand{\barPCall}[1]{\overline{\PCall}_{#1}}

\title{Auto-encoding GPS data to reveal individual and collective behaviour}

\author[1,2]{Saint-Clair Chabert-Liddell \orcidlink{0000-0001-5604-7308}}
\author[2]{Nicolas Bez \orcidlink{0000-0002-3526-671X}}
\author[3]{Pierre Gloaguen \orcidlink{0000-0003-2239-5413}}
\author[1]{Sophie Donnet \orcidlink{0000-0003-4370-7316}}
\author[2]{Stéphanie Mahévas \orcidlink{0000-0001-7707-2502}}

\affil[1]{Université Paris-Saclay, Agroparistech, INRAE, UMR MIA Paris-Saclay, 91120 Palaiseau, France}
\affil[2]{Marbec, Univ Montpellier, CNRS, Ifremer, IRD, Sète, France}
\affil[3]{Université Bretagne Sud, UMR CNRS 6205, LMBA, F-56000 Vannes, France}
\date{}
\begin{document}

\maketitle

\begin{abstract}
We propose an innovative and generic methodology to analyse individual and collective behaviour through individual trajectory data. The work is motivated by the analysis of GPS trajectories of fishing vessels collected from regulatory tracking data in the context of marine biodiversity conservation and ecosystem-based fisheries management. 
 We build a low-dimensional latent representation of trajectories using convolutional neural networks as non-linear mapping. This is done by training a conditional variational auto-encoder taking into account covariates. The posterior distributions of the latent representations can be linked to the characteristics of the actual trajectories. 
 The latent distributions of the trajectories are compared with the Bhattacharyya coefficient, which is well-suited for comparing distributions. Using this coefficient, we analyse the variation of the individual behaviour of each vessel during time. For collective behaviour analysis, we build proximity graphs and use an extension of the stochastic block model for multiple networks. This model results in a clustering of the individuals based on their set of trajectories. 
 The application to French fishing vessels enables us to obtain groups of vessels whose individual and collective behaviours exhibit spatio-temporal patterns over the period 2014-2018.
\end{abstract} 

\section{Introduction}
\label{sec:introduction}

The conservation and sustainable use of marine resources is a world priority that requires an accurate assessment of the status of exploited marine populations. Two sources of information are key to estimating the status of exploited marine populations around the world, namely scientific surveys and commercial fisheries' statistics. Both sources provide a partial picture of the reality. For instance, scientific surveys, when they exist, only occur at some specific time of the year and miss seasonal issues. Commercial fisheries statistics are provided all year long by the fishermen, but their use in the stock estimation process often assumes that fishing data are independent. However, fishermen collaborate or compete to increase their yields, so that the above independence assumption does not hold, with possible, but not well anticipated, detrimental effects on the assessment of the status of the exploited marine populations \citep{poos2007experiment, fulton2011human}. 
This assumption has recently been investigated. \cite{joo2021} established interaction networks between fishing vessels, studying pairs of their trajectories (called dyads) with proximity and coordination indicators  \citep[see][for a review of indicators]{joo2018}. Using the graphlet correlation distance, \cite{roux2023} tested whether their corresponding interaction networks were compatible with the assumption that the trajectories were independent or not. However, the authors acknowledge that a weakness of these approaches lies in the subjective and limited choices of indicators used to describe dyads and measure pairwise distance.

The general motivation of the present work is therefore to develop a method for analysing trajectories that does not require defining indicators for dyads prior to the analysis, but that statistically reveals relevant metrics for comparing trajectories. Trajectories are characterized by a two-dimensional time series (geographical positions at each point in time). We propose to learn from the observed trajectories a low-dimensional space in which some sub-regions can be related to particular behaviours. Once projected in this low-dimensional space, trajectories can then be compared using any relevant metric tools (distance, means, clusters\dots). These comparisons may subsequently allow us to derive groups of vessels sharing similar behaviours. \\

To learn a low-dimensional \emph{representation}  of the trajectories \citep{bengio2013representation},  we propose to use a probabilistic hierarchical model. We assume that the trajectories follow a probabilistic distribution whose parameters arise from a low dimensional latent space through a non-linear transformation (mapping), called \emph{decoder}. 
The prior distributions in the latent space are assumed to be known,  and the resulting posterior distributions can then be recovered by learning an inverse mapping.  
In practice, we use a variational approach \citep{blei2017variational} to approximate the posterior distribution. 
The variational distribution is restricted to the Gaussian family, whose parameters are assumed to be non-linear functions of the observations. The mapping from observations to the parameters of the variational distribution is called the \emph{encoder}. 
Both the decoder and the encoder are parameterized by neural networks and their parameters are learnt jointly by maximizing a proxy of the log-likelihood.

This modelling framework is known as a variational auto-encoder \citep[$\VAE$;][]{kingma2014, rezende2014stochastic} and can be seen as non-linear extensions of probabilistic PCA \citep{bishop1998bayesian, lawrence2005probabilistic}. 
Compared to a classical auto-encoder \citep{kramer1991nonlinear}, the agreement between the variational distribution and the prior on the latent space acts as a regularization term, which prevents over-fitting and brings more structured latent representations. 
Adding covariates to the model is straightforward and leads to a conditional variational auto-encoder \citep[$\CVAE$;][]{kingma2014}, which ensures that the latent representations do not reflect spurious correlation with these known covariates.

The choice of a neural network architecture for both mappings is guided by the nature of the data (spatio-temporal) and by the question (exploration of the trajectory as a whole, i.e. large- and medium-scale properties). In this work, we specifically use Convolutional Neural Networks (CNN). The interested reader may refer to recent reviews on deep generative models and neural network architectures for spatio-temporal data \citep{gao2022generative,graser2023}.

Quantifying similarity between latent representations of trajectories is usually done through the computation of indices using point estimates (e.g. Euclidean distance or cosine similarity on the mean of the variational distributions). We chose here to exploit the full benefits of the probabilistic trajectory modelling, and we quantify the similarity between latent representations with the Bhattacharyya Coefficient \citep[$\BC$;][]{bhattacharyya1943measure}, a measure of overlap between probability distributions. Using the full distribution allows capturing the uncertainty associated with the latent representation of each data trajectory. The overlap coefficient can be interpreted as a quantification of similarity between daily individual spatio-temporal behaviours.

Comparison of trajectories using similarity between their representations in the low-dimensional latent space is useful in several ways. First, comparing the trajectories of a single fishing vessel for different fishing days (adjacent or not) helps to quantify the variation of its daily individual behaviour. Alternatively, for a given fishing day, if we compare the trajectories of a group of vessels we can derive a proximity network, i.e. a binary graph where a node is a vessel and an edge between two vessels denotes a similarity above a pre-defined threshold.    
Using an extension of the Stochastic Block Model \citep[SBM; ][for a review]{holland1983, lee2019review} for collections of networks \citep{chabertliddell2023}, these daily networks can be analysed to derive groups of vessels with similar behaviours, to understand how the different groups relate to each other and how they change over time periods.

The general approach outlined in this article and specifically, the concept of encoding individual trajectories to later achieve the clustering of individuals, could be of interest in analysing trajectories of different types of objects (or individuals). The method is therefore presented in a general context, and details specific to our case study are set and discussed in the applications and results.

The manuscript is organised as follows. Section \ref{sec:material-and-methods} describes the trajectory data, introduces the conditional variational auto-encoder ($\CVAE$) model used for analysis, and presents the tools employed for analysing the latent representations. Then the results are divided into several subsections. Section \ref{sec:choose-ls} covers the selection of the latent space for representation, and Section \ref{sec:latent-space} provides intuitive insights about these representations. Section \ref{sec:individual-behaviours} deals with the analysis of individual behaviours, while Section \ref{sec:prox-graph} focuses on collective behaviours. In this latter section, we present the construction of proximity graphs from the latent representations of fishing trajectories and the application of network analysis tools for grouping fishing vessels. 
A discussion on the application of our method to fishery science case study and some perspectives conclude the paper in Section \ref{sec:discussion}.

\section{Material and methods}
\label{sec:material-and-methods}
 \subsection{Data set}
\label{sec:data-set}

The data consists of $M=20512$ daily GPS trajectories from $B = 33$ different fishing vessels over a 5-year period. The number of trajectories per vessel ranges from $140$ to $968$ (median $701$).
All the trajectories start and end at the same point, a fishing harbour in French Brittany. 
The positions of the vessels are obtained from the mandatory Vessel Monitoring System \citep[VMS, ][]{hinz2013}, recording at a roughly hourly rate. 
To synchronize all the trajectories (i.e., have exactly the same timestamp), we use a linear interpolation of the
position recorded by the VMS as recommended by \cite{joo2018}.
Hence, each trajectory is a matrix of dimension $d_{\traj} = H \times D$. In this study, the number of hours is $H = 24$ and we consider 2d trajectories (geographical positions expressed in longitude and latitude), so $D = 2$.

We denote $\trajobs_m \in \mathbb{R}^{d_\traj}$ the $m^{th}$ trajectory $(1 \leq m \leq M = 20512)$, and $\trajsetobs = \lbrace \trajobs_{m} \rbrace_{m = 1,\dots, M}$ the complete data set.
In the following, when necessary, we will use the notation $\trajobs_{(b,t)}$ to specifically refer to the trajectory of the vessel $b$ at day $t$, if it exists, where $1\leq b \leq B$, and $t$ spans over the whole period of the data set. 
In a general setting, these trajectory data can be enriched with additional covariates. 
For each trajectory $1\leq m \leq M$, 
we denote $\covar_{m} \in \mathbb{R}^{d_\covar}$ the set of available covariates. 
In this paper, the available covariate is the day of the year at which the trajectory occurs. For $j\in \llbracket 1;365 \rrbracket$, the time of the year is given by the vector $\left(\cos \left( \frac{2\pi\times j}{365} \right), \sin \left( \frac{2\pi\times j}{365} \right)\right)$ (and consequently ${d_\covar}=2$) to take into account the periodicity.

\subsection{Methods for learning the latent representations}

\subsubsection{Probabilistic generative model}

We consider a probabilistic framework where the observed trajectories are realizations of random vectors $\trajset = \{\traj_m\}_{m = 1, \dots, M}$ following a probabilistic distribution characterized by a highly non-linear transformation of the realization of low dimension latent random vectors. 
Formally, in this paper, we suppose that:
\begin{equation}
\begin{array}{rl}
    \latent_{m} & \overset{\text{ind.}}{\sim} p_{0} := \mathcal{N}\left(0, I_{d_\latent} \right),~1\leq m \leq M  \\
    \traj_{m} \vert \latent_{m} = \latentobs_{m}; \covar_{m} & \overset{\text{ind.}}{\sim} \mathcal{N}\left(\decod(\latentobs_{m}, \covar_{m}; \decodpar); \sigma^2 I_{d_\traj}\right)
\end{array}
\label{eq:generative-model}
\end{equation}
where 
\begin{itemize}
    \item $\mathcal{N}(\mu, \Sigma)$ denotes the normal distribution with mean $\mu$ and matrix of variance $\Sigma$;
    \item $I_d$ denotes the identity matrix of dimension $d$;
    \item $\decod(\cdot, \cdot; \decodpar)$ is a function from $\mathbb{R}^{d_\latent}\times \mathbb{R}^{d_\covar}$ to $\mathbb{R}^{d_\traj}$, depending on some unknown parameter $\decodpar \in \mathbb{R}^{d_\decodpar}$;
    \item $\sigma^2 > 0$ is a variance parameter whose role is discussed in Section \ref{rem:variance}.
\end{itemize} 

The key part of this generative model is the function $\decod(\cdot, \cdot; \decodpar)$, which maps the latent space to the space of trajectories. 
This function, often called \textit{decoder}, must be chosen within a class of non-linear mappings, parameterised by $\decodpar$, typically neural networks. 
The precise choice for this function is discussed in Section \ref{sec:cnn}. %\todo{ADD REF}.
 The main learning objective is then to estimate $\decodpar$ from observed $\trajsetobs$.

\subsubsection{Variational inference} \label{sec:inference}

For the probabilistic latent model given by Equation \eqref{eq:generative-model}, the log-likelihood is given by integrating over the whole latent space, namely:%\footnote{In the following, $p(U)$ is a generic and overloaded notation for the probability density function of a random variable $U$. When this p.d.f. depends on the value of some variable $V$, it will be denoted by $p(U\vert V)$ when $V$ is random and $p(U;V)$ when $V$ is not random.}:

\begin{equation*}
    \log p(\trajsetobs ; \covarset, \decodpar) = \log \int p(\trajsetobs | \latentsetobs; \covarset, \decodpar)p_{0}(\latentsetobs) \text{d}\latentsetobs.
\end{equation*}
The integral on the right-hand side has no explicit expression, and is usually hard to approximate by numerical approximations or Monte Carlo methods, due to the dimension of $\latentsetobs$. 
An alternative inference method is the variational approach \citep{blei2017variational}. 
This approach exploits a surrogate of the likelihood function, the evidence lower bound (ELBO).
For any probability distribution function $q(\latentsetobs \vert \trajsetobs ; \covarset,\varpar)$, parameterised by some parameter $\varpar \in \mathbb{R}^{d_\phi}$:
\begin{eqnarray}
    \log p(\trajsetobs ; \covarset, \decodpar) &\geq&  \log p(\trajsetobs ; \covarset, \decodpar) - \DKL(q(\cdot\vert \trajsetobs ;\covarset,\phi)\| p(\cdot \vert \trajsetobs ; \covarset, \decodpar)) \label{eq:kl-elbo}\\
    &=& \underbrace{\mathbb{E}_{q(\cdot \vert \trajsetobs ; \covarset, \phi)}\left[\log p(\trajsetobs \vert \latentset ; \covarset, \decodpar) \right] 
    - \DKL\big(q(\cdot \vert \trajsetobs; \covarset, \phi)\| p_{0}(\cdot)\big)}_{{=: \text{ELBO}(\decodpar, \varpar)}}
    \label{eq:elbo},
\end{eqnarray}
where $\DKL$ stands for the Kullback-Leibler divergence. 
The inequality becomes an equality in the single case where $q(\cdot \vert \trajsetobs; \covarset,\varpar) = p(\cdot \vert \trajsetobs; \covarset, \decodpar)$, and maximising the ELBO with respect to $\varpar$ (for a fixed $\decodpar$) is equivalent to minimising the Kullback-Leibler divergence between $q(\cdot \vert \trajsetobs ; \covarset, \varpar)$ and $p(\cdot  \vert \trajsetobs; \covarset, \decodpar)$ as seen in Equation \eqref{eq:kl-elbo}. 
Therefore, by maximising the ELBO jointly with respect to both $\decodpar$ and $\varpar$, for a given family of variational distribution $q(\cdot \vert \trajsetobs; \covarset,\varpar)$, one obtains $\widehat{\decodpar}$ that maximises a proxy of the log-likelihood and $\widehat{\varpar}$ that gives the best member of the family chosen to approximate the distribution $p(\cdot \vert \trajsetobs; \covarset, \widehat{\decodpar})$.

A flexible and standard framework for the variational family is to assume normality, i.e., that $q(\latentobs_{m} \vert \trajobs_{m}; \covar_{m}, \varpar)$ is the p.d.f. of a normal distribution in dimension $d_\latent$, with mean $\varmean_{m}\in \mathbb{R}^{d_\latent}$ such that $\varmean_{m} = \varmean(\trajobs_{m}, \covar_{m}; \varpar)$ and a diagonal covariance matrix whose diagonal is noted $\varvar_{m} = \varvar(\trajobs_{m}, \covar_{m}; \varpar)$.    Both means and variances, therefore, depend on observed data through a mapping parameterised by $\varpar$.
    This mapping $h(\cdot, \cdot; \varpar): (\trajobs_m, \covar_m) \mapsto (\varmean_m, \log\varvar_m)$ is usually called the \textit{encoder}. 

The loss function to minimise is $-ELBO$, taking its form in Equation \eqref{eq:elbo}, comprises two terms whose expressions in the Gaussian cases come easily: a reconstruction error term,
\begin{equation}\label{eq:reconstruction-error}
    - \mathbb{E}_{q(\cdot \vert \trajsetobs; \covarset,\phi)}\left[\log p(\trajsetobs \vert \latentset ; \covarset, \decodpar) \right] = \frac{1}{2\sigma^2}\sum_{m=1}^{M}\mathbb{E}_{q(\cdot ; \phi)}\left[\|\trajobs_{m} - g(\latent_m, \covar_m; \decodpar) \|_{2}^{2} \right] +\frac{Md_{\trajsetobs}}{2}\log{{\sigma^2}} + \mathcal{O}(1), 
\end{equation}
and a regularization term which takes advantage of the fact that the dimensions in the latent space are independent and that the divergence is the sum of the divergence by dimension,  
\begin{equation}\label{eq:regularisation}
  \DKL(q(\cdot \vert \trajsetobs;\covarset,\phi)\| p_{0}(\cdot)) = \frac{1}{2}\sum_{m=1}^{M}\sum_{k=1}^{d_\latent} \bigl(\varvar_{k} + \varmean^{2}_{k} - \log \varvar_{k} - 1\bigr).
\end{equation}

The minimization of the negative ELBO is then done through a stochastic gradient descent algorithm with mini-batches using the ADAM optimizer \citep{kingma2015adam}. 
During the optimization,  for gradient back-propagation purposes we use the reparameterization trick \citep{kingma2014, rezende2014stochastic}, each entry of the encoder $\latentobs_{m}$ is simulated independently as $\latentobs_{m} = \varsd_{m} \epsilon + \varmean_{m}$ where $\epsilon$ is the realization of a  $\mathcal{N}(0,I_{d_\latent})$ (cf. Figure \ref{fig:archi_cvae}). 

In accordance with the deep learning literature, the model with covariates is called a conditional variational auto-encoder ($\CVAE$) and the one without covariates, where $\encod$ and $\decod$ are not functions of $\covar$,  a variational auto-encoder ($\VAE$).

\paragraph{\textbf{Remark about the variance parameter $\sigma^2$}}\label{rem:variance} 
In the context of variational inference, the variance parameter $\sigma^2$ is difficult to estimate and is usually fixed \citep{skafte2019reliable}. If fixed, for a Gaussian distribution, choosing the variance corresponds to choosing the importance of the regularization term compared to the reconstruction term \citep{higgins2017betavae}. Higher variance leads to higher regularization, as can be seen in Equation \eqref{eq:reconstruction-error}. When $\sigma^2$ is fixed to $1$, the reconstruction term of the negative ELBO corresponds to the L2 loss.    %When willing to quantify uncertainty, one may use quantile regression \citep{rodrigues2020beyond}.

\subsubsection{Neural network architecture}\label{sec:cnn}

The practical implementation of the framework described above needs a parametric family of non-linear mapping for the encoder and the decoder.
This is done by choosing appropriate neural network architectures. 

It is classic to view a trajectory as an image on a two-dimensional grid at the cost of losing the time stamps in the trajectories  \citep{olive2020deep, liang2021unsupervised}. In the present study, the temporal dimension of the trajectory is important for assessing the synchronization or time lag of collective behaviour. So as not to lose it, we use the time stamps of the trajectory as pixels and the spatial position as channels. 

\begin{figure}[t]
    \centering
    \includegraphics[width=\mywidth]{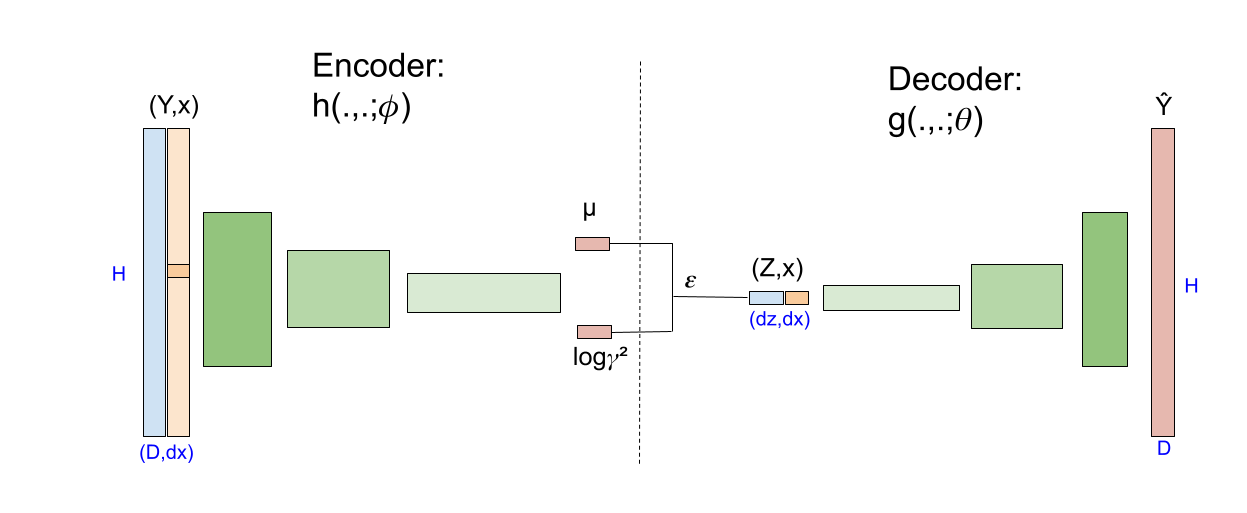}
    \caption{Model and its neural network architectures. Each data is transformed into an encoder input, the covariates are duplicated to match the length of the 2-dimensional trajectory. For each box, the vertical axis is ordered (image pixels) while the horizontal axis is unordered (channels). The input of each neural network is in blue, the covariates in orange, the 3 hidden layers in green and the output in red. }
    \label{fig:archi_cvae}
\end{figure}

A $2$-dimensional trajectory of length $H$ can be viewed as a one-dimensional image of $H$ pixels with $2$ channels. 
Using this representation, a common architecture used in deep generative models is a convolutional neural network (CNN). For generating trajectories, we build from the architecture proposed by \citet{radford2016} and successfully used by \citet{roy2022} to generate bird foraging trajectories. This architecture was adapted to a lower dimensional problem and to handle data with covariates. The two covariates $\covar$ are added to the trajectory image on two additional channels and replicated to match the length of the trajectory. 
The architecture of the encoder is then chosen to be as symmetric as possible with respect to the decoder. 
A synthetic view of the model and its architecture is shown in Figure \ref{fig:archi_cvae}. 
More details about the architecture and pseudocode for \texttt{PyTorch} implementation \citep{pytorch} are provided in Appendix \ref{app:sec:cnn}.

\subsection{Methods for the analysis of the latent representations}

\subsubsection{Quantifying similarities in the latent
space}\label{sec:latent-distance}

Each trajectory $\trajobs_{m}$, possibly  knowing its date $\covar_m$, is mapped through the encoder $\encod(\cdot, \cdot; \hat{\varpar})$ to a  Gaussian distribution with mean $\hat{\varmean}_{m}$ and variance $\hat{\gamma}^{2}_{m} I_{d_\latent}$ (therefore assuming independence). 
In this latent space, the similarity between two encoded trajectories is quantified by a measure of overlap between distributions, the Bhattacharyya coefficient \citep[$\BC$;][also referred to as the Hellinger affinity]{bhattacharyya1943measure}, which integrates the entire variational distributions and has an expression in closed-form for independent multivariate Gaussian distributions. 
The $\BC$ between two fishing vessels $b$ and $b'$ fishing on day $t$ and $t'$, 
is given by:

\begin{equation}
    BC_{b,b'}^{t,t'} = %BC(Z^{i,t}, Z^{j,t'}) = 
    \int_{z} \sqrt{q(z \vert \trajobs_{(b,t)};\covar_{(b',t')}, \hat{\varpar}) \cdot q(z \vert \trajobs_{(b',t')}; \covar_{(b',t')}, \hat{\varpar})}\text{d}z.
\end{equation}
$BC_{bb'}^{tt'} \in (0,1]$ and is equal to one if and only if
the two distributions are the same. The closed-form
expression of the $\BC$ between two multivariate Gaussian distributions with diagonal covariance matrix is
the following:

\begin{equation*}
    \BC_{b,b'}^{t,t'} = \prod_{k=1}^{d_\latent}\sqrt{\frac{2\hat{\varsd}_{(b,t), k}\hat{\varsd}_{(b',t'), k}}{\hat{\varsd}^{2}_{(b,t), k} + \hat{\varsd}^{2}_{(b',t'), k}}} \exp{\biggl (-\frac{1}{4}\frac{(\hat{\varmean}_{(b,t), k} - \hat{\varmean}_{(b,t), k})^2}{\hat{\varsd}^{2}_{(b,t), k} + \hat{\varsd}^{2}_{(b',t'), k}}\biggr)}.
\end{equation*}

Analysis of the similarity between trajectories is then performed at the individual level by looking at the overlap coefficient of a vessel $b$ at two different times $t$ and $t'$, noted $\BC_{b}^{t,t'} := \BC_{b, b}^{t,t'}$, or at the collective level by looking at the overlap of two vessels $b$ and $b'$ on the same day $t$, noted $\BC_{b, b'}^{t} := \BC_{b, b'}^{t,t}$.
The overlap coefficient is further binarized to look for high overlapping trajectory representations, by computing $\BC^{t,t'}_{b,b'} \geq s$ where $s \in (0,1]$ is a users' defined threshold (set to $0.8$ in the present study). Finally, the behaviour stability index of a vessel over a set of couples of days $S$ is computed as the mean of the binarized coefficients:

\begin{equation}\label{eq:pc}
    \PC_{b}^{S} =  \frac{1}{\vert S\vert }\sum_{(t,t') \in S}\mathbf{1}_{\BC_{b}^{t,t'}\geq s}.
\end{equation}
Similarly, within the fleet, we define a general stability index  computed over all possible trajectories irrespective of the vessel to serve as a baseline for comparing the stability of individual behaviours: 
\begin{equation}\label{eq:pcall}
    \PC^{S} =  \frac{2}{B(B-1)}\sum_{b, b'= 1, b\neq b'}^{B}\frac{1}{\vert S\vert}\sum_{(t,t') \in S}\mathbf{1}_{\BC_{b,b'}^{t,t'}\geq s},
\end{equation}

$\PCall_b = \PC_{b}^{S_g}$, where $S_g = \{(t,t'): t \neq t'\}$ indicates the proportion of similar trajectories per vessel globally over time (global stability). Respectively, $\PCall = \PC^{S_g}$ quantifies the global stability of the fleet.  \\

 $\PClag_b = \PC_{b}^{S_d}$, where $S_d = \{(t,t'): t' =  t+1\}$, gives information on the proportion of similar trajectories from one day to the next (daily stability), while $\PClag = \PC^{S_d}$ measures the daily stability within the fleet.

\subsubsection{Tools for analysing proximity graphs}\label{sec:tools-graph}
A collection of proximity graphs is built to analyse the collective behaviour. During a time period, two vessels are said to be in proximity if the distributions of their trajectories in the latent space for the same day are highly overlapped, i.e.
if the $q$-quantile of their $\BC$ distribution for
the time period is greater than a given threshold $s$. That is, for each time period $S \subset \llbracket1;T\rrbracket$, the adjacency matrix of the proximity graph is defined as:

\begin{equation}\label{eq:graph}
  A^{S}_{bb'} = \begin{cases} 
  1 & \mbox{ if } \frac{1}{\vert S\vert}\underset{t\in S}{\sum} \mathbf{1}_{BC_{b,b'}^{t} \geq s} \geq q, \\
  0 & \mbox { otherwise. }
  \end{cases}    
\end{equation}
Only vessels with observed trajectories during this time period are represented in the graph. Self-interactions and interaction between vessels with no common fishing days are encoded as \texttt{NA}.

An extension of the Stochastic Block Model \citep[SBM, ][]{holland1983} is then used to analyse the proximity graphs, called networks hereafter. 
SBM allows grouping fishing vessels exhibiting similar connectivity patterns with the rest of the fleet. Fishing vessels from the same group are said to be equivalent. Hence, each network is described in a condensed way by its mesoscale structure (the proportion of each group and the probability of connection within and between each group). Extension of the SBM for collection of networks \citep[colSBM, ][]{chabertliddell2023}, potentially enables finding a common mesoscale structure describing all the networks of the collection, and to determine if this structure is sufficient to describe the whole collection (if not, the collection is divided into sub-collections of networks with similar mesoscale structure). This model does not take into account the vessel ID. Mathematical details of the model are provided in Appendix \ref{app:sec:sbm}.\\

\begin{figure}[t]
    \centering
    \includegraphics[width=\textwidth]{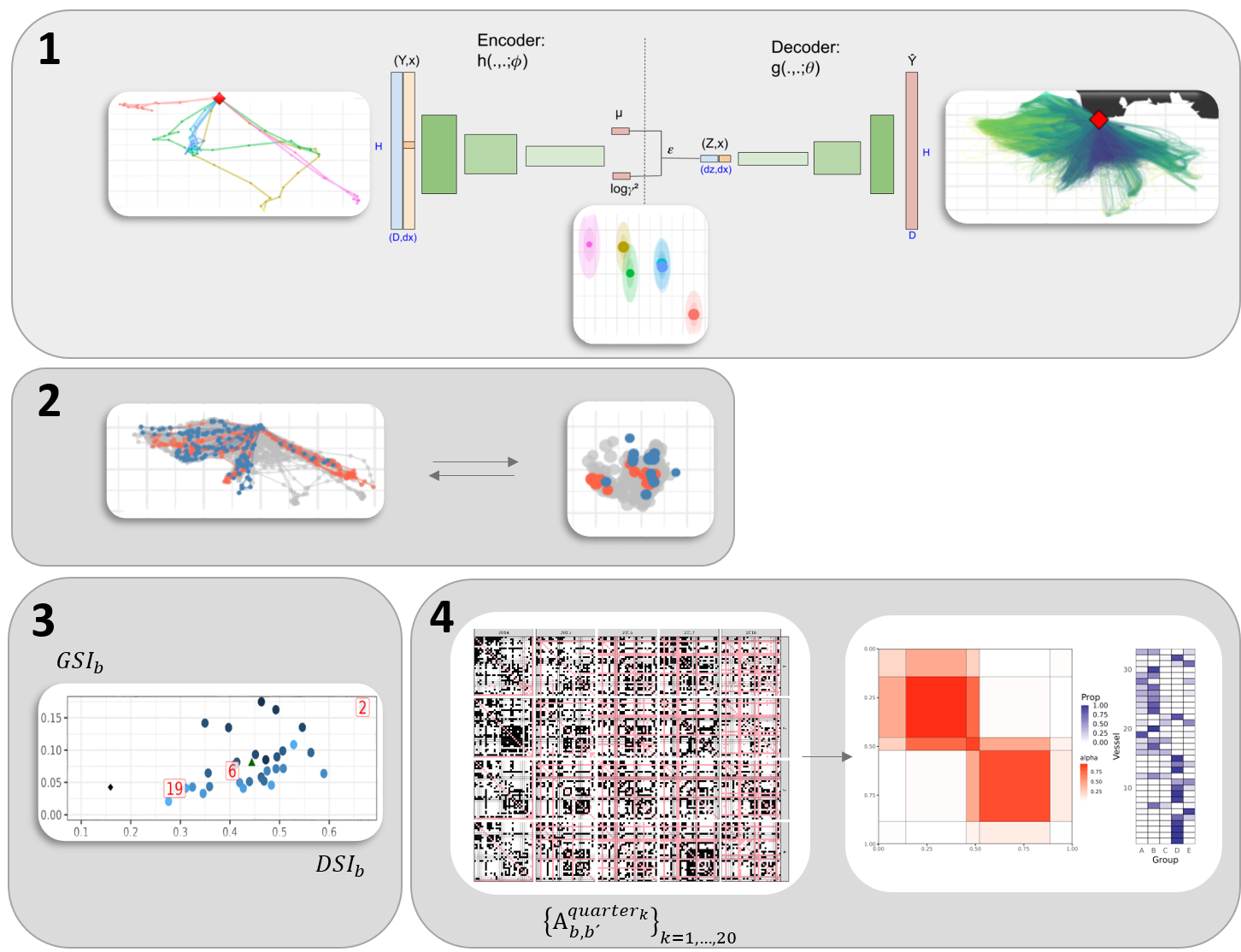}
    \caption{Workflow. $\mathbf{1}$) Learning latent representations of trajectory data using (conditional) variational auto-encoders. $\mathbf{2}$) Analysis of the relation between the latent representations and the trajectory data. $\mathbf{3}$) Analysis of daily and global stability of individual behaviours using the Bhattacharyya Coefficients computed from the latent representations. $\mathbf{4}$) Analysis of the collective behaviour with the Stochastic Block Model using proximity graphs between individuals built from the Bhattacharyya Coefficients computed from the latent representations.}
    \label{fig:workflow}
\end{figure}

A workflow of the data analysis, explaining how the different methods introduced in this section interact, is given in Figure \ref{fig:workflow}.

\section{Application and results}
\label{sec:application-and-results}

\subsection{Choosing the latent space}\label{sec:choose-ls}

\subsubsection{Optimal dimensions of the latent space}

\begin{figure}
    \centering \includegraphics[width = \mywidth]{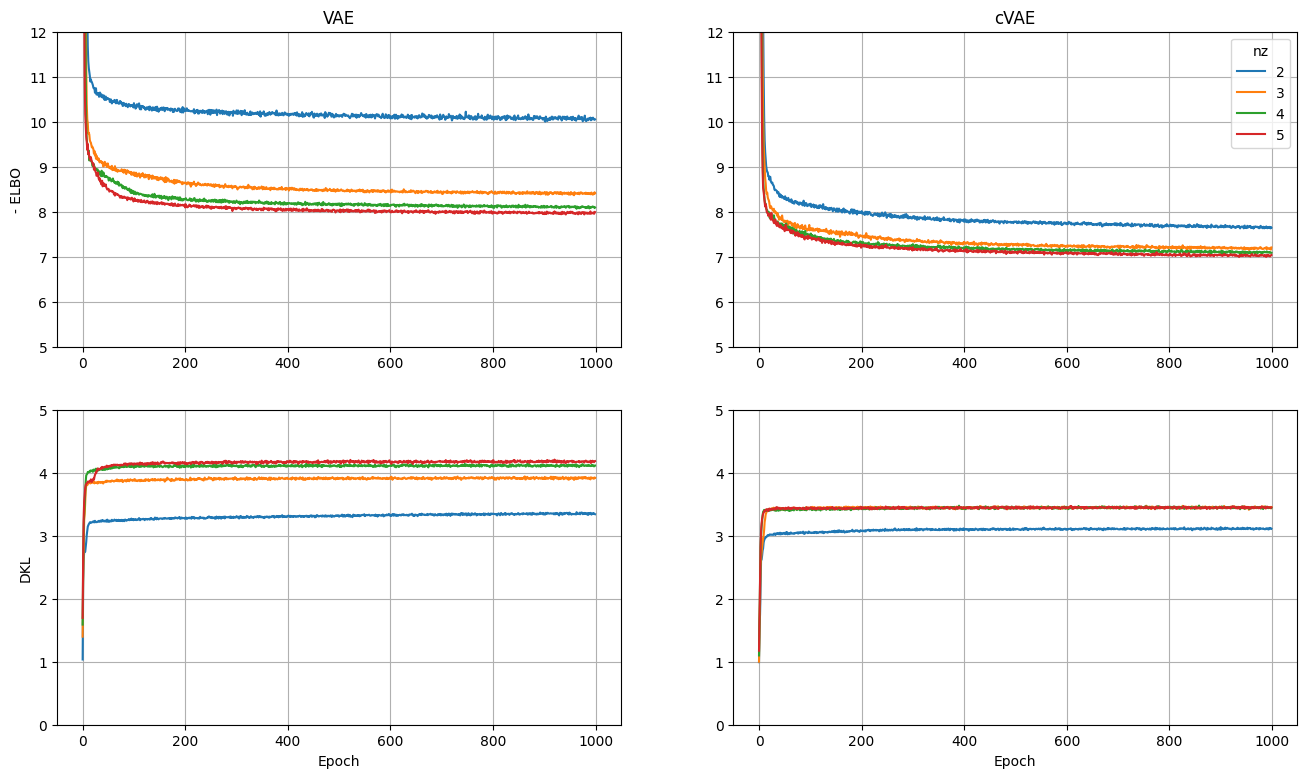}
    \caption{Loss (-ELBO) and regularization term (DKL) for the first 1000 epochs for the VAE and the CVAE with 2 to 5 latent dimensions.} 
    \label{fig:elbo}
\end{figure}

The aim is to choose the model with the smallest-dimensional latent space that is sufficient to describe the trajectory data.
Figure \ref{fig:elbo}  shows the loss ($- \ELBO$) and the regularization term ($D_{KL}$) for the VAE and the CVAE with $d_\latent = 2, \dots, 5$. From $d_\latent = 3$ for the CVAE and $d_\latent = 4$ for the VAE, the loss does not decrease much with additional dimensions, and the $D_{KL}$ term stops increasing meaning that the structure of the latent space does not depart from the prior distribution anymore, and hence adding dimensions to the latent space is not necessary to describe the trajectories.
From now on, we consider $d_{\latent} = 3$ for the CVAE model and $d_{\latent} = 4$ for the VAE model.

\subsubsection{Ordering the dimensions of the latent space}

\begin{figure}
\centering \includegraphics[width = \mywidth]{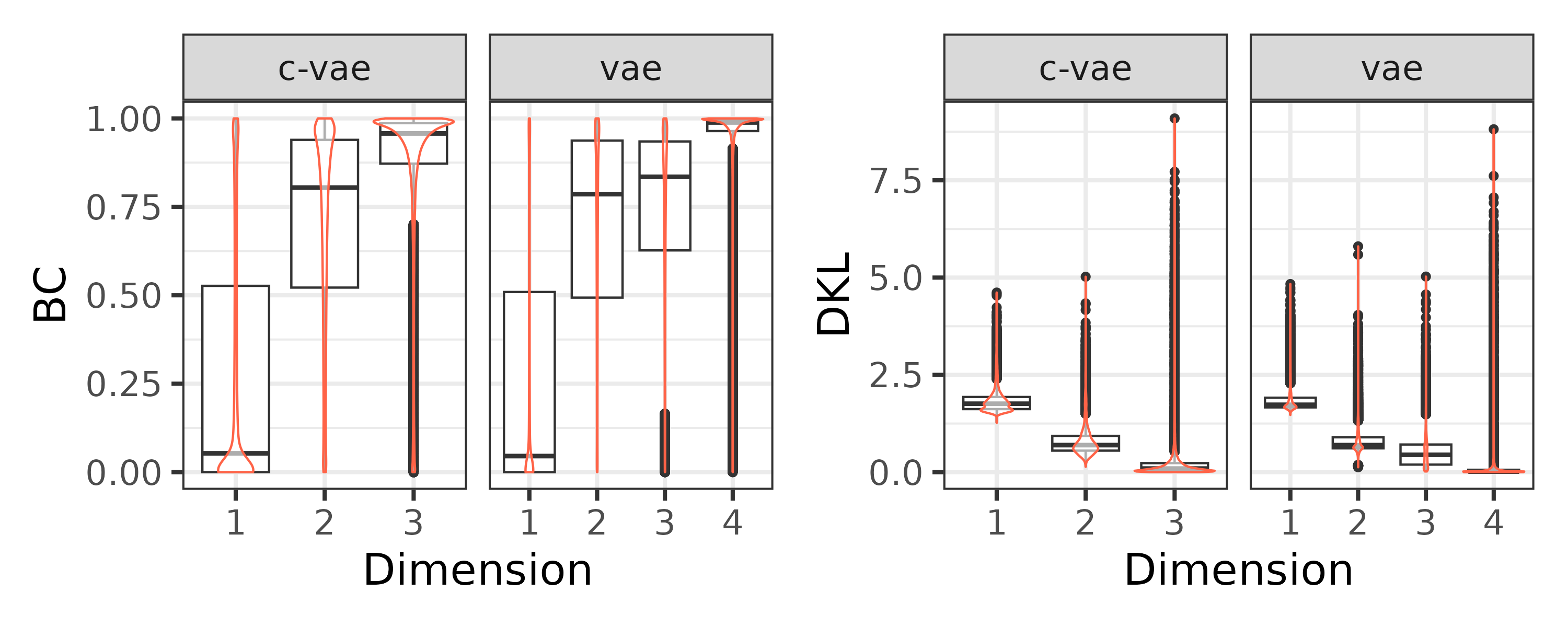}
\caption{\label{fig:dimension}Empirical distribution of BC between
couples of variational distribution and Kullback-Leibler divergence
between variational distribution and normal Gaussian for each dimension
of the latent space.}
\end{figure}

As the dimensions of the variational distribution are independent (diagonal covariance matrix), the $\BC$ between two variational distributions is the product of the BC between each dimension of the two distributions. We can then look at the distribution of the $\BC$s for each dimension and order the dimensions of the latent by level of importance (Figure \ref{fig:dimension}, left). If $\BC$s are close to one, it means that the distributions have a very high overlap on this dimension and that, this dimension has a low impact on the total metric and does not bring much information on differentiating the trajectories in the latent space. 

The dimensions where $\BC$ are close to $1$ correspond also to the dimensions where the variational distributions are close to the prior ($\DKL$ massively close to 0; Figure \ref{fig:dimension}, right). These dimensions are thus only useful to describe a few unusual trajectories, which correspond to the one in the tail of the distribution for both $\BC$ and $\DKL$. It happens that while the analysis of the outputs of CVAE (resp. VAE) relies on three (resp. four) dimensions, the first two (resp. three) dimensions are sufficient to represent the trajectories in the latent space without significant loss of information.

\subsection{Linking latent representations to
trajectories}\label{sec:latent-space}

With the objective to explore how representations in the latent space are related to trajectories in the geographical space and to choose between VAE and CVAE frameworks, 3 vessels with very contrasting individual behaviours have been selected (vessels number $2$, $6$ and $19$). 

\subsubsection{Choosing between CVAE and VAE}

\begin{figure}
\centering \includegraphics[width = \mywidth]{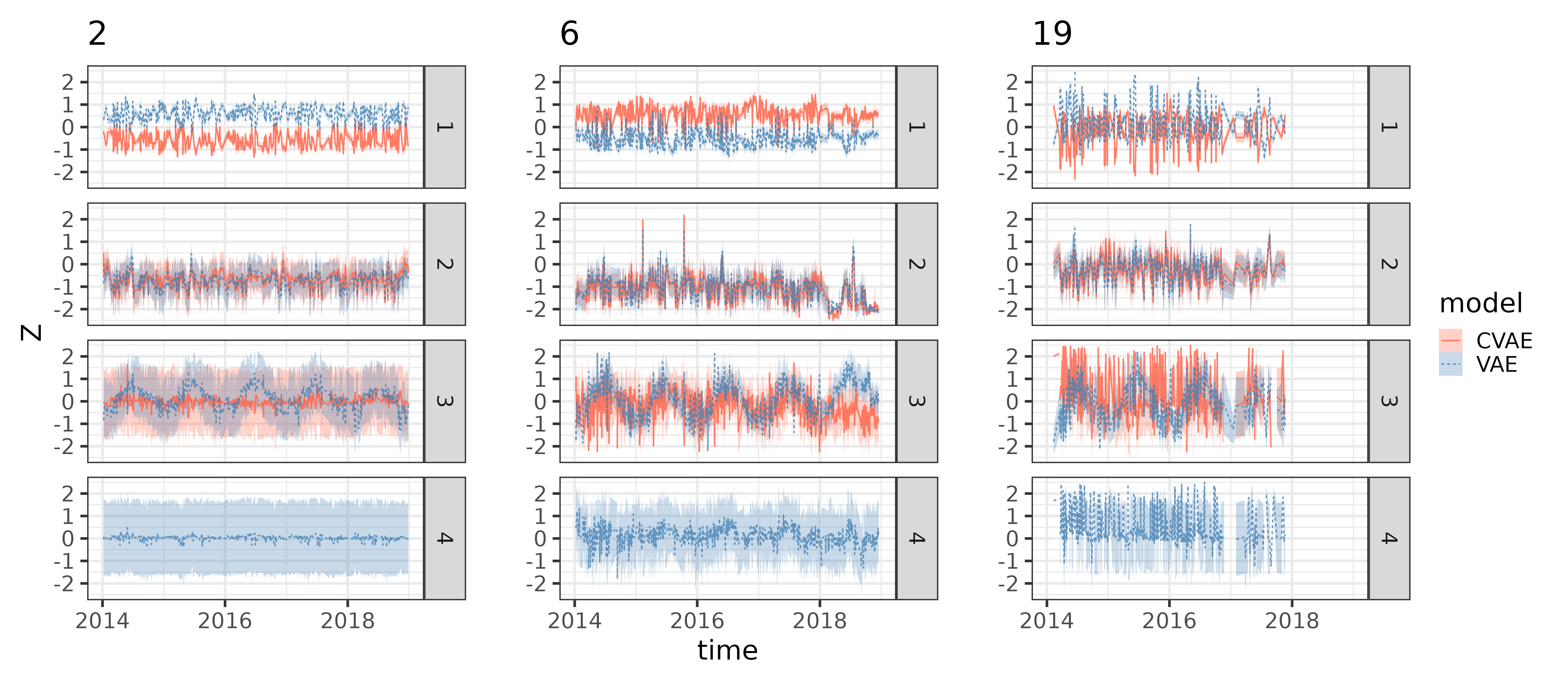}
\caption{Time series of the latent variational distributions represented by the curves $t \mapsto ,\mu_{(b, t),k} $ and $t \mapsto  \mu_{(b, t),k} \pm 1.96\gamma_{(b, t),k}$ for vessels $b = 2,6,19$ and  $k = 1, \dots, d_\latent$.}
\label{fig:ts-mu}
\end{figure}

\begin{figure}
{\centering \includegraphics[width = \mywidth]{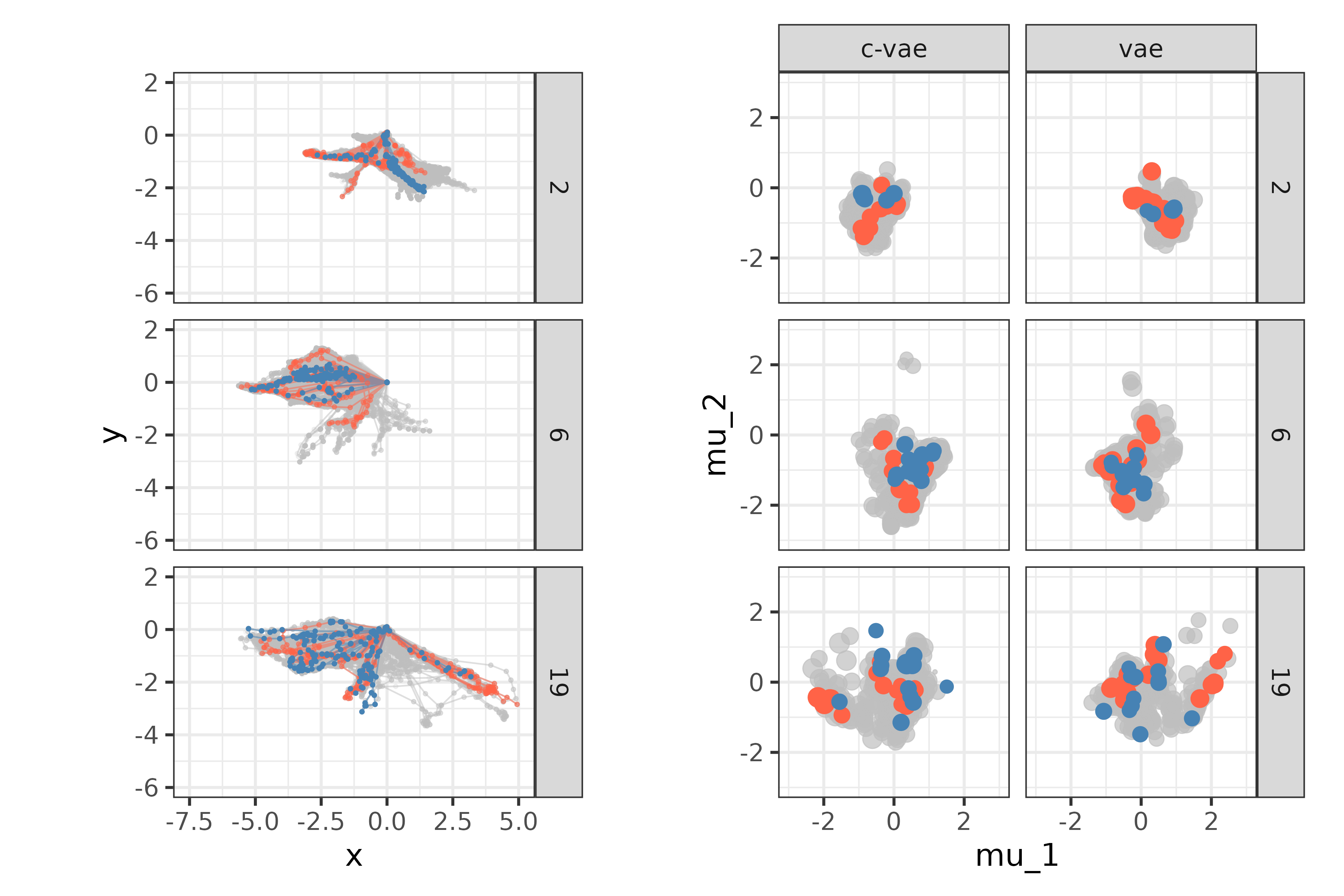}
}
\caption{Trajectories and latent space for fishing vessels $2$, $6$ and $19$. The highlighted trajectories and points correspond to the one of June 2015 in orange and December 2015 in blue. The $x$-axis, $y$-axis and size of points in the latent space correspond respectively to $\mu_{(\cdot, \cdot),1}$, $\mu_{(\cdot, \cdot),2}$ and $\mu_{(\cdot, \cdot),3}$.}
\label{fig:traj-ls-3boat}
\end{figure}

For each of the three selected vessels, we investigate the time series of their variational distributions on each dimension of the latent space for both the VAE and the CVAE models (Figure \ref{fig:ts-mu}), their trajectories and the corresponding plots of the mean values in the first three dimensions of the latent space (Figure \ref{fig:traj-ls-3boat}). 
 
For the VAE model, $\mu_{(b,\cdot),3}$ exhibits a very clear sinusoidal seasonality (Figure \ref{fig:ts-mu}), taking negative values in winter (around $-2$) and positive values in summer (around $2$). This effect is taken into account by the covariates of the CVAE for which times series do not show any seasonal variations anymore. 
In this regard, a CVAE model with a three-dimensional latent space and time as covariate provides equivalent outputs than a VAE approach with four dimensions. Given that the third dimension of the CVAE is practically useless in terms of graphical representations, this is a first indication for promoting CVAE against VAE.\\

Both the trajectories and the mode of their latent representations 
are progressively less compact for vessel $2$, $6$ and $19$ (Figure \ref{fig:ts-mu} and grey points in Figure \ref{fig:traj-ls-3boat}) indicating that trajectories with increasing length (i.e. trajectories whose point furthest from the port is getting farther and farther away) are mapped by expanding points in the latent space for both CVAE and VAE.  
Vessel $2$ exhibits only two kinds of behaviours in its trajectories in December, one sailing west and the other one sailing southeast. The fishing behaviours in June are more diverse but one trajectory going west overlaps the trajectory that occurred in December while being slightly longer. These two particular trajectories are overlapped in the CVAE latent space near the origin, while they correspond to two different points in VAE. 
A similar pattern emerges for vessel $19$, the points corresponding to the long trajectories to the south-east direction in June and December overlap for the CVAE (near $(-1.5,0)$) but not for the VAE (near $(2,0)$).

If the same intrinsic fishing behaviour produces longer trajectories in summer than in winter due to a change in the distribution of the target species, then a CVAE approach that translates these two different trajectories into a single point in latent space provides a better mapping of behaviours than a VAE approach does. Hereafter, we will thus only consider the latent space of the CVAE model.\\

\subsubsection{Deeper analysis of the latent space for CVAE}
\begin{figure}
    \centering
    \includegraphics[width = \mywidth]{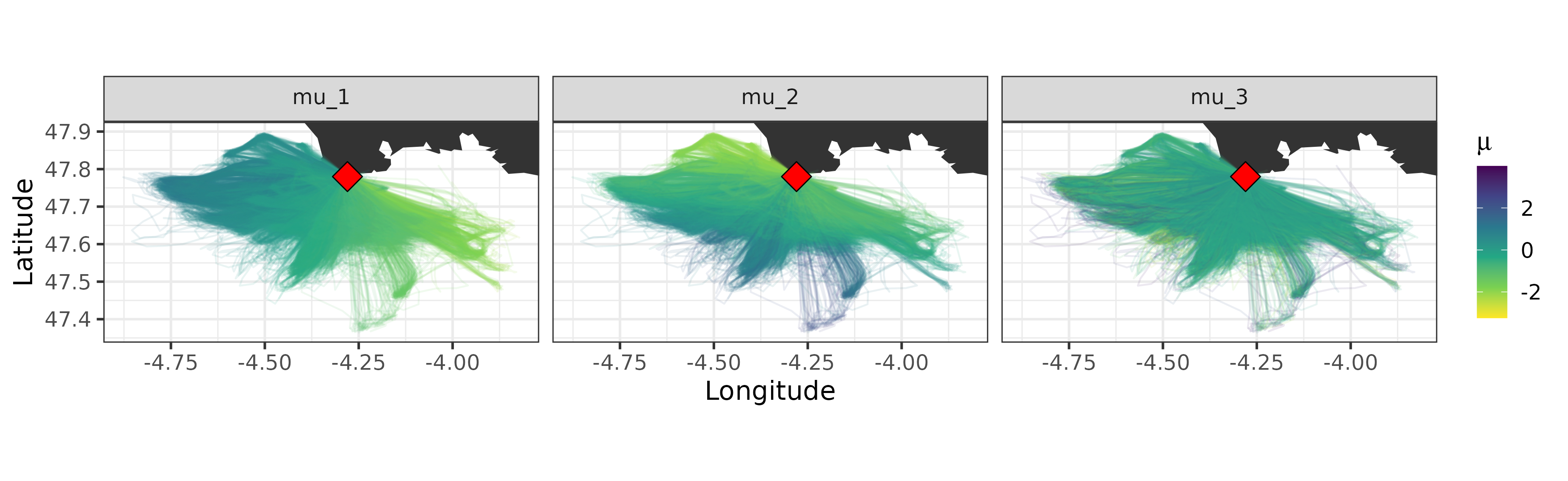} 
    \includegraphics[width = \mywidth]{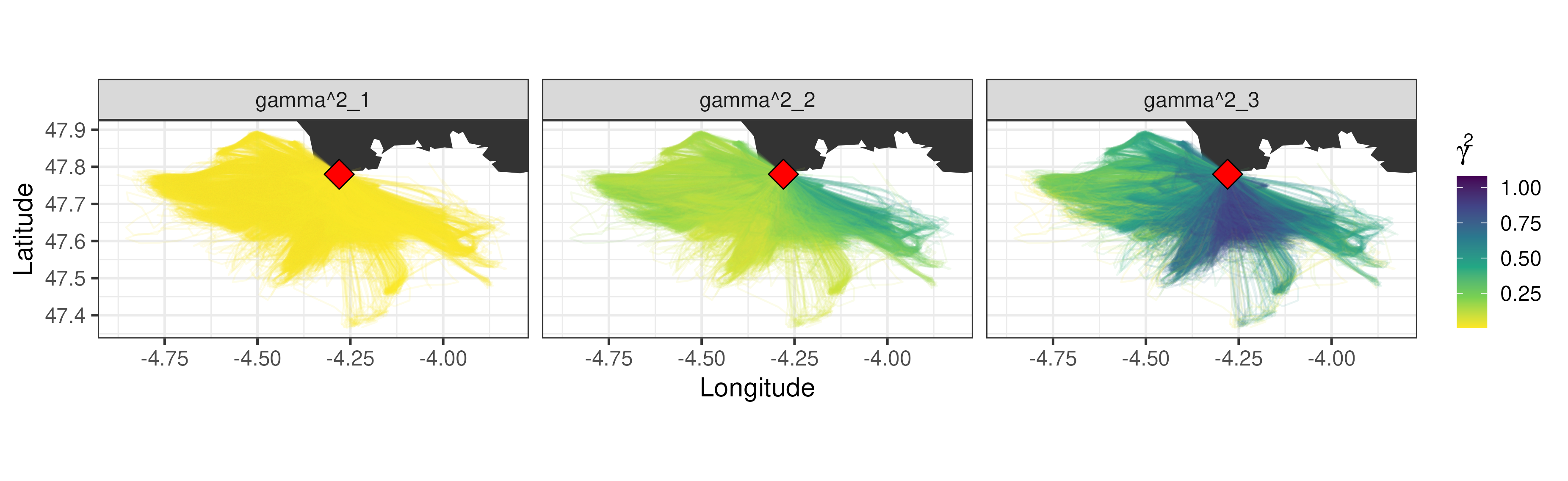}
    \caption{Trajectories coloured by the mode of the parameters of their variational distribution in the latent space ($\varmean$ above and $\varvar$ below) on each of the $3$ dimensions of the CVAE. The red lozenge represents the harbour (geographical position of vessel departure and arrival), which is located at $(-4.28, 47.78)$.}
    \label{fig:traj_mu}
\end{figure}

We show how the dimensions of the latent space are related to the length of the trajectories and to an east-west and north-south gradient.
For each trajectory data, we look at the parameters $\varmean$ and $\varvar$ of its variational distribution from the CVAE model. In Figure \ref{fig:traj_mu} each trajectory of the dataset (before normalization) is coloured according to $\varmean$ and $\varvar$ for each dimension.  Interpretation for each parameter follows:
\begin{enumerate}
    \item $\varmean_1$ varies on a longitudinal gradient with $\varmean_1$ ranging from smallest (negative) values  ($\varmean_{1} \approx -2$) to largest positive values ($\varmean_{1} \approx 1$) from east to west, with $\varmean_1 = 0$ corresponding to trajectories going south-west from the harbour. 
    \item $\varmean_2$ varies along a latitudinal gradient, with trajectories from north to west corresponding to the smallest values ($\varmean_2 \approx -2$), while trajectories far to the south have the largest values ($\varmean_2 > 2$).
    \item $\varmean_3$ does not exhibit a clear spatial pattern, except from some trajectories going west with the largest values ($\varmean_3 > 2$).
    \item $\varvar_1$ is very low for each trajectory, indicating good confidence in the ability to characterize the direction of the trajectory using the first dimension of the latent space.
    \item $\varvar_2$ is the smallest for the trajectory going south and the largest for the easternmost trajectories. The second dimension of the latent space will have difficulty distinguishing the latitudinal position of trajectories in the eastern part of the region.  It corresponds to an area with a low density of observations.
    \item $\varvar_3$ is the largest (close to one) for short trajectories going south of the harbour. Those trajectories have also $\varmean_3 \approx 0$ meaning that the variational distribution did not depart from the prior for these trajectories, and this dimension is not informative. The longest trajectories to the west have much lower variance in their variational distribution. Their corresponding means are $\varmean_1 \approx 1$ and $\varmean_3 > 2$.
\end{enumerate}

\begin{figure}
    \centering
\includegraphics[width = \mywidth]{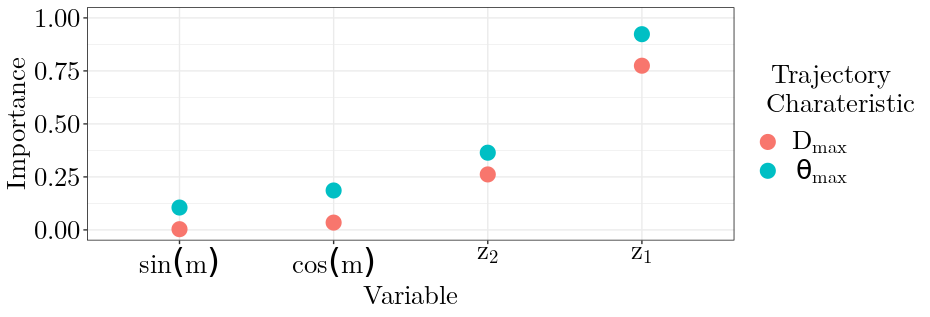}
    \caption{Sensitivity of trajectory characteristics ($D_{\max}$: distance to the port; $\theta_{\max}$: main direction) to the conditioning variable and the first 2 dimensions.}
    \label{fig:SI}
\end{figure}

To further quantify the relationship between the latent representations and the actual trajectories, we evaluate the sensitivity of two key trajectory characteristics, i.e. the maximal distance to the harbour (denoted $D_{\text{max}}$) and the main direction of the trajectory (denoted $theta_{\text{max}}$, the angle corresponding to the maximally distant point in the trajectory), with respect to the two first dimensions ($\latentobs_1$ and $\latentobs_2$) of the latent space and to the month of the year (coded as $\cos(m)$, $\sin(m)$). 
To this end, a grid with $\latentobs_1$ spanning from $-2$ to $1.2$,  $\latentobs_2$  from $-2$ to $2$ and $m$ corresponding to each month of the year is built ($\latentobs_3$ is fixed to $0$). 
Trajectories are simulated by using the decoder $\decod(\cdot,\cdot;\widehat{\decodpar})$ obtained after training the CVAE model.
From these trajectories, the response variables $D_{\text{max}}$ and $\theta_{\text{max}}$ are computed. 
Then, these two quantities are predicted using a random forest \citep[with the \texttt{ranger} \texttt{R} package;][]{ranger} having $(\latentobs_1, \latentobs_2, \cos(m)$, $\sin(m)$) as input. 
The estimated sensitivity index is computed using the Sobol MDA index, designed to be a proxy of the total Sobol index for random forests accounting for dependency between variables \citep{benard2022mean}. 
This index ranges from $0$ to $1$, equalling $0$ if there is independence between the predictor and the response, and $1$ if the response is a deterministic function of the predictor. 
Figure \ref{fig:SI} displays these indices for the four explanatory variables. As expected, one can see that the maximal distance and main directions are strongly sensitive to the first dimension, and, to a lesser extent, to the second dimension. 
We also can see that the distance to the port is not sensitive to the time of the year, but the direction is affected by this covariate.

In summary, the latent space allows to reliably discriminate trajectories along a positive east-west gradient on the first dimension, northward trajectories with low values on the second dimension and long trajectories with high values on the third dimension. On the other hand, short and/or southward trajectories are more difficult to identify in latent space.

\subsection{Exploring individual behaviours}\label{sec:individual-behaviours}

\begin{figure}
\centering \includegraphics[width = \mywidth]{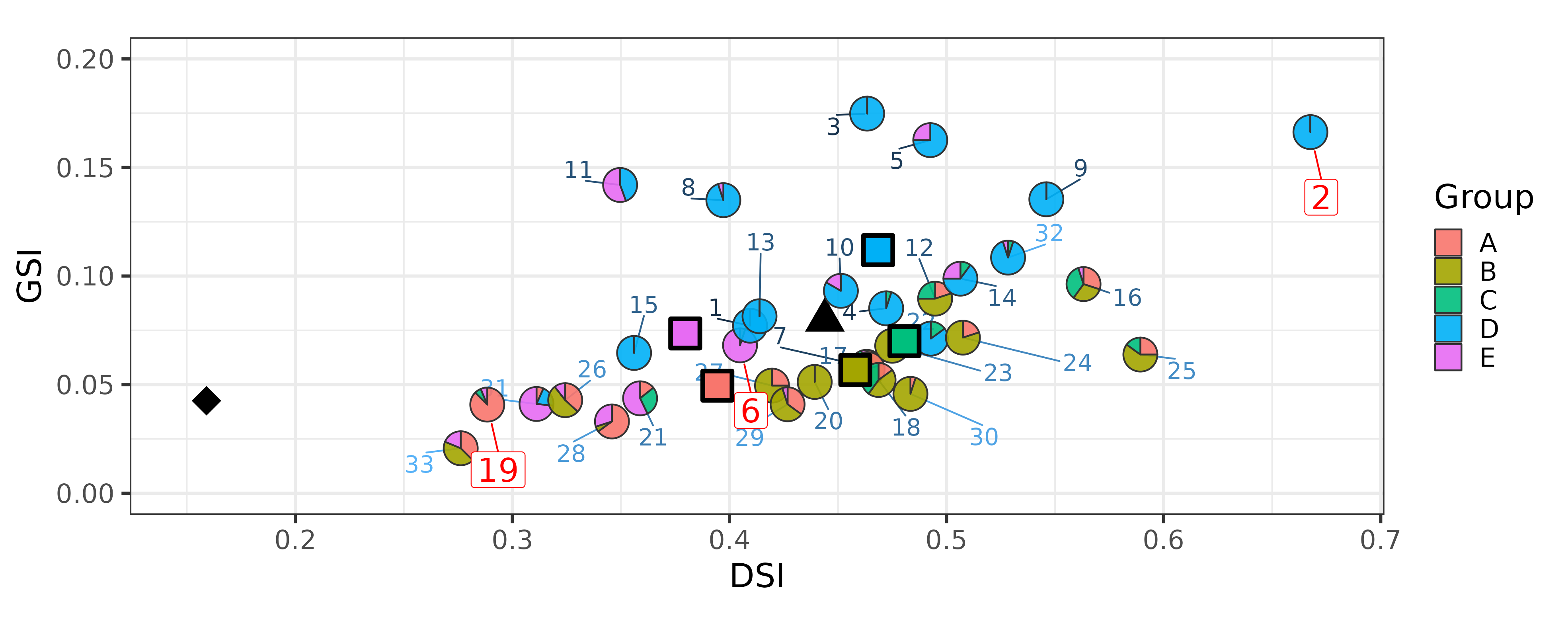}
\caption{Proportions of trajectories with high
overlap between their latent representations ($\BC \geq 0.8$) for the $33$  fishing vessels. On the x-axis, the proportion for adjacent days trajectories ($\PClag_b$) and on the y-axis, the proportions for all couples of trajectories ($\PCall_b$). Each vessel (each point) is identified by its number and the proportion of group memberships found by the colSBM (refer to the right of Figure \ref{fig:block-colsbm}) given by the pie-chart colours. The three contrasted vessels $(2, 6, 19)$ used to illustrate the exploration of individual behaviour are written and framed in red. The black diamond shows the proportions computed for the whole fleet ($\PClag$ and $\PCall$), while the black triangle represents the average of individual stabilities. Averages of individual stabilities for each group are shown in black-boxed squares.}\label{fig:bc-nav}
\end{figure}

Using the individual stability index and the global stability index at the fleet scale, we investigate the behaviour of the vessels. For all fishing vessels $\PClag_{b} > \PCall_{b}$ (Figure \ref{fig:bc-nav}) indicating that fishermen tend to reproduce similar trajectories from one day to the next and don't change drastically fishing grounds every day. 
The two indices are correlated, but heterogeneity can be found in the variety of individual behaviours between vessels. For instance, vessels $2$, $3$ and $5$ have high global stability, but vessel $2$ has by far the most stable trajectories between following days ($\PClag_{2} > 0.65$). Vessels have very different daily stabilities, ranging from $0.27$ to $0.67$, with a fleet average of about $0.45$.
In comparison,  the global stability of each vessel is very low (ranging from $0.025$ to $0.175$) indicating that vessels have many kinds of individual behaviour, i.e. they globally often change of trajectories.\\

As expected, amongst the three selected vessels mentioned in section \ref{sec:latent-space}, the vessel with the most variable time series (vessel $19$ in Figure \ref{fig:ts-mu}) is also the one with the smallest temporal stabilities ($\PClag_{19}$ and $\PCall_{19}$). It explores widely the fishing area and its trajectories change frequently. Conversely, the one with the most stable time series and compact latent space  (vessel $2$ in Figure \ref{fig:traj-ls-3boat}) is the one with the largest global and daily stability indices (see Figure \ref{fig:bc-nav}). The behaviour of vessel $6$ lies in between and is close to the average individual stability (green triangle in Figure \ref{fig:bc-nav}).

At the fleet level, the vessels have a very low daily stability (about $0.16$, black diamond at the bottom left of Figure \ref{fig:bc-nav}). 
Each of the vessels has on average more day-to-day stable trajectories than the fleet  ($\PClag_{b} > \PClag$). This is not always true for distant time trajectories, as about a third of the fleet has $\PCall_{b} < \PCall$. This indicates that vessels have many different behaviours that are explored below using network analyses.

\subsection{Exploring collective behaviour}\label{sec:prox-graph}

Choosing adequate time spans to analyse collective behaviour is of primary importance. Relying on the CVAE outputs, a daily time step is likely to result in highly unstable behaviour. Conversely, considering the entire period ($5$ years) masks seasonal behaviour or yearly trends. The quarter is a common time period used to define fishing behaviour \citep{ices2022working}, and therefore we build 20 (one per quarter over 5 years) proximity graphs from the latent representations of the vessels' trajectories using the Bhattacharyya coefficients (see section \ref{sec:tools-graph}). The adjacency matrices of the proximity graphs are shown in Appendix \ref{app:sec:proxy-graph}, Figure \ref{fig:proxy-graph-trim}. 

\subsubsection{Clustering the vessels into groups}
We show that the structure of the proximity graphs is stable over time and that the fleet contains two main groups of vessels exhibiting exclusive spatio-temporal behaviours. 

\begin{figure}
\centering 
\includegraphics[width = \mywidth]{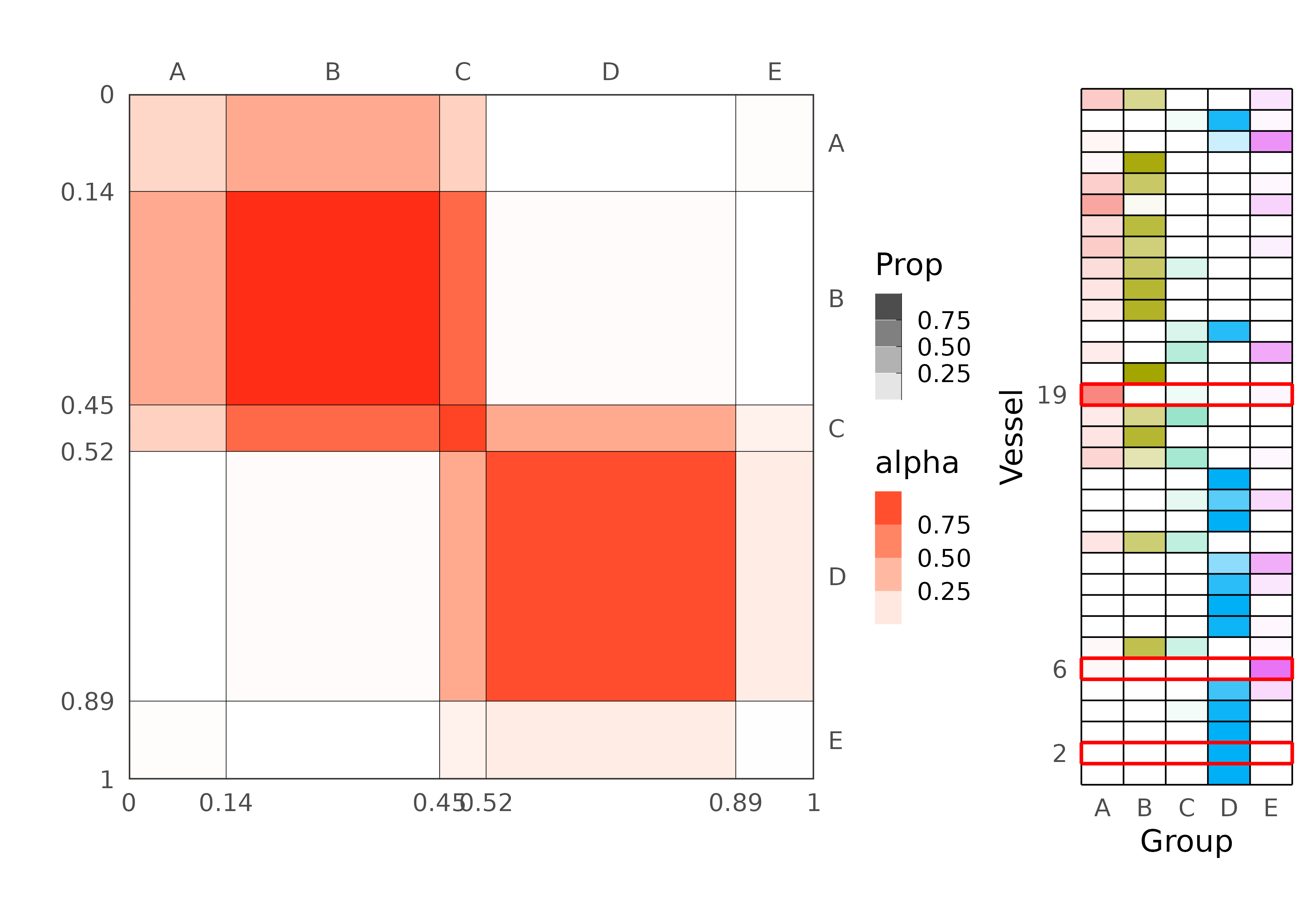}
\caption{\label{fig:block-colsbm}Left: Common mesoscale structure of the collection. The
dark lines separate the $5$ groups, the size of the blocks stands for the block proportions, while the shade of reds (alpha) gives the average number of connections between and within groups.
Right: Group membership proportions (Prop) found by the 
colSBM model for each fishing vessel over the collection of 20 quarterly proximity graphs. The three selected vessels are highlighted (vessels 2, 6 and 19).}
\end{figure}

To analyse the networks, we use the colSBM model introduced in Section \ref{sec:tools-graph}. 
The model selection criterion does not favour the partition of the collection of $20$ proximity networks into sub-collections. It means that the structures of the networks are similar between quarters and that the common mesoscale structure shown in Figure \ref{fig:block-colsbm} is sufficient to describe the 20 quarterly proximity graph.
The fleet is divided into $5$ groups of fishing vessels labelled $A$ to $E$ (Figure \ref{fig:block-colsbm}, right) and ordered in the matrix summary of the mesoscale structure from top to bottom, left to right (Fig. \ref{fig:block-colsbm}, left). Groups $B$ and $D$ are two exclusive communities of highly internally connected fishing vessels, representing about a third of the fleet each. $A$ and $E$ are smaller groups that are on the periphery of respectively groups $B$ and $D$, with group $E$ regrouping vessels nearly without mutual connection.
It is worth noticing that there are almost no connections between groups $A$, $B$ and groups $D$, $E$.% ($(A \cup B) \cap (D \cup E) = \varnothing$). 
 Finally, $C$ is a small group of fishing vessels with heterogeneous behaviours and connecting with both sides of the fleet. 
%On the left of Figure \ref{fig:block-colsbm}, we show the block proportions for each vessel across the $20$ networks. 
The memberships of fishing vessels to group $D$ are very stable across quarters, with however ephemeral switch to the peripheral group $E$ (Figure \ref{fig:block-colsbm}). On the contrary, vessels from the rest of the fleet switch more often between groups $C$, $B$ and $A$.\\  

The analysis of the fishing behaviour in terms of daily and global stability can be further highlighted by the group memberships (Figure \ref{fig:bc-nav}). %(Figure \ref{fig:bc-colsbm}). 
Groups $B$ and $D$ representing two-thirds of the fleet have on average the same daily stabilities of their trajectories ($\barPClag{B} = 0.46$ and $\barPClag{D} = 0.47$), with strong variabilities around. However, fishing vessels from group $B$ have nearly systematically more diverse trajectories, and lower global stability, than the ones from group $D$ ($\barPCall{B} = 0.06 < \barPCall{D} = 0.11$).
Fishing vessels in the peripheral groups ($A$ and $E$) have much lower daily stabilities ($\barPClag{A} = 0.38$ and $\barPClag{E} = 0.37$) and behave alike group $B$ for global stability.

\subsubsection{Analysis of the trajectories and their latent representations by groups}
\begin{figure}
    \centering
    \subfloat[][]{
            \includegraphics[width = \mywidth]{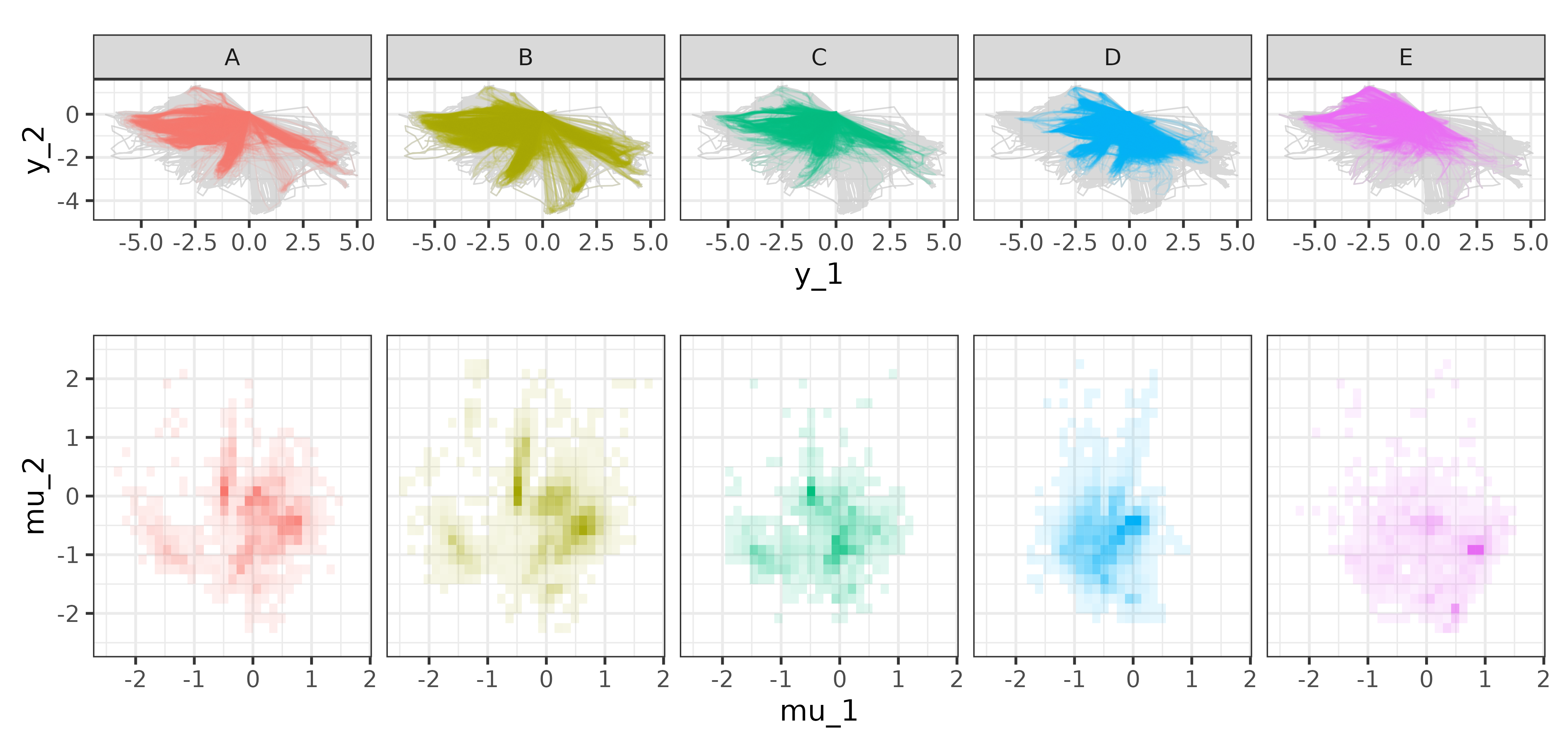}\label{fig:traj-ls-group}}\\
    \subfloat[][]{
\includegraphics[width = \mywidth]{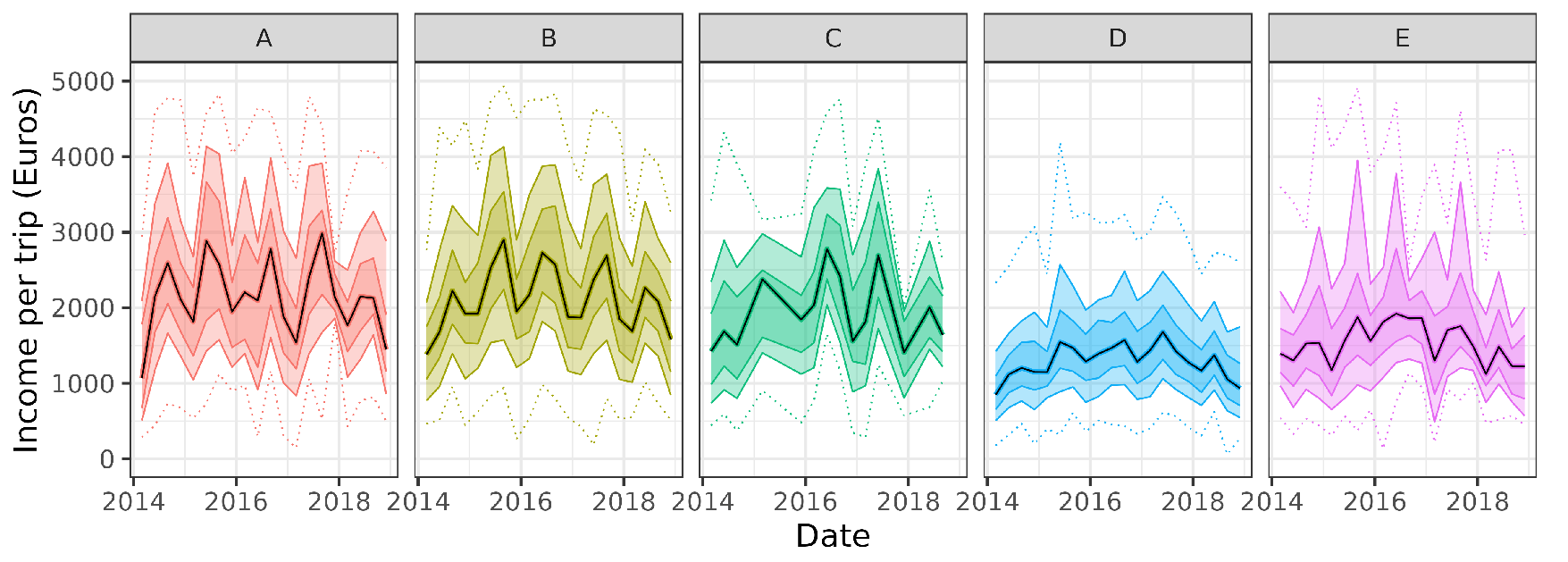}\label{fig:income-group}}
    \caption{(a) Top: trajectories ordered by the groups found with the colSBM model. Bottom: distribution of the mean of the latent representations $(\varmean_{(\cdot, \cdot), 1}, \varmean_{(\cdot, \cdot), 2})$ for the same sets of trajectories. 
    (b) Quantile distribution of the income per fishing day by quarter and by the group found with the colSBM model. The black line is the median, while the shades depict the $\alpha$ and $1-\alpha$  quantiles  for $\alpha \in \{.01, .1, .25\}$.}
    
\end{figure}

Yet, we can investigate the specificities of trajectories within groups. 
The distributions of the mean of the first two dimensions of the latent representations of the trajectories for the five groups are shown in the lower part of Figure  \ref{fig:traj-ls-group}, while the trajectories are shown in the upper part. Both are very different for groups $D$ and $E$ from the ones of groups $A$, $B$ and $C$. We can briefly summarize the trajectory behaviour for each group as follows:
\begin{enumerate}
    \item [$A, B$]  Latent representations of trajectories of these two groups look very similar. Both have a wide range of values in the latent space, with high concentrations around the same place. There are some differences between the two groups. Group $B$ takes more extreme values in the latent space (for instance on the top left, which corresponds to the long trajectory to the south). But, as seen in Figure \ref{fig:block-colsbm}, both groups are composed of the same fishing vessels. In particular, they make long trajectories. 
    \item [$C$] Looks very similar to group $A$ on the trajectory space, but the concentration is more uniform in the latent space.
    \item [$D$] Values for the first dimension are taken on a short interval from $-1$ to $0.5$, with a high concentration around $(0,0)$. Notice how the distribution is almost complementary from the one of group $B$ on the latent space, and included on the trajectory space. The fishing vessels of this group make shorter trajectories.
    \item [$E$] Takes lower values in the second dimensions and higher in the first dimensions than the other groups. Its fishing vessels have the northernmost trajectories in the trajectory space. They are the only ones to consistently explore this fishing ground.
\end{enumerate}

In the initial dataset, we had catches in values (euros) for each daily trajectory (fishing trip).  We investigated whether these groups differed in terms of gross revenue. The main source of gross value for the fleet consists of fishing crustaceans (\textit{Nephrops norvegicus}), but there are side significant incomes which come from fishes (mainly monkfish (Lophius), hake (merluccius merluccius), sole (solea solea), megrim (Lepidorhombus), and john dory (Zeus faber)). 
The gross values per fishing trip are variable between the years and also between vessels within the group.  They are significantly different between groups $A$, $B$, $C$ and groups $D$, $E$ ($1400$€ on average for group $D$ and $2400$€ on average for groups $B$ and $A$; Figure \ref{fig:income-group}). 
The vessels from group $E$ have median incomes close to the one of group $D$, but in some quarters have much higher incomes in the upper quantiles.

As the trajectories of groups $D$ and $E$ are less distant from the port and therefore generate lower fuel costs, we may wonder whether these vessels are not looking for a cost-revenue trade-off. It would be interesting to explore net revenues, which unfortunately were not available for this study.

\section{Discussion and perspectives}\label{sec:discussion}

We proposed the utilization of deep learning methods, especially variational auto-encoders, to learn a low-dimensional representation (a probability distribution) of trajectory data. This approach allowed for the transformation of complex trajectory data into a more interpretable representation facilitating the analysis of the individual and collective spatio-temporal behaviours. Temporal variability in the latent representations was attenuated by using the day of the year of a trajectory as a covariate in the variational auto-encoder. This statistical conditioning also allowed to decrease the number of dimensions of the latent space from $4$ to $3$. 
In our study, we used an overlap index, the Batthacharyya Coefficient, between the distributions of the representations to quantify the similarities in individual behaviour and build proximity graphs. These proximity graphs facilitated the clustering of fishing vessels, resulting in groups of vessels based on the stability of individual behaviours. The exploration of these groups has highlighted differences in spatial and economic patterns. It was observed that groups exhibiting shorter trajectories have lower gross values per trip, most likely revealing a cost/revenue trade-off strategy.  Additionally, these groups exhibited lower overall effort (cost) per trip, suggesting a more efficient utilization of resources.

In fisheries, one of the aims of building groups of vessels with similar behaviour is to identify reference groups (reference fleets) in order to estimate indices of abundance from their catches, which reveal inter-annual variations in fish abundance. The quality of these abundance indices depends on good sampling by fishermen of the species' distribution area. One possibility is to use these groups to construct a stratified estimator that takes into account the spatial structure of exploration by vessels to remove bias in current estimators and yet improve stock assessment and management decisions \citep{poos2007experiment, alglave2022combining}.

Whether these indices can be improved by this method or not relies on the assumption of trajectory independence \citep{bez2003catchability}. We could use the generative aspect of the CVAE model to generate trajectories to test a null hypothesis of trajectory independence. This aspect of the model is not present in this paper, except to relate the characteristics of the trajectories to the dimensions of the latent space.

We now discuss the choice of the neural network architectures, the indices to calculate proximity between latent representations and the hypotheses that were made when comparing behaviours and building proximity networks.

In this work, we used Convolutional Neural Networks (CNN) as the neural network architecture for both the encoder and the decoder. 
An alternative would have been to use Recurrent Neural Networks, in particular a Long-Short Term Memory (LSTM) architecture, as they are classic for temporal data. However, \citet{roy2022} point to the inability of LSTM in a generative model to consistently generate trajectories which start and end at the same position (cycle), which is a common feature in all the trajectories of our data set.  In addition, LTSM are particularly appropriate to i) capture short-scale behaviour while we are more interested in medium and large-scale characteristics of the trajectory and ii) provide an operational prediction of a vessel's next position based on its past trajectory \citep{capobianco2021}, for instance, to avoid collisions \citep{jia2022motion}.

Because of the heavy parametrization of the nonlinear mapping learnt by the VAE, the embedding usually lacks interpretability. In order to circumvent this drawback, some authors make a tradeoff between reconstruction error and interpretability by still using a non-linear encoder but a linear decoder \citep[e.g.]{svensson2020interpretable}, while others aim at disentangling the dimension of the latent space to represent features  \citep[e.g.]{higgins2017betavae}, as it is difficult to relate a dimension to a feature of the data. 
The trajectory of a fishing vessel being smooth and regular, starting and ending at the same point in space, we were able to obtain very low-dimensional representations. This allowed for post hoc analysis to interpret the latent representations and relate them to the trajectories. Low-dimensional space was also beneficial for graphical representation as by properly ordering the dimension of the latent space, we did not need to rely on further dimensional reduction tools \citep[e.g. t-sne]{van2008visualizing} to visualize the latent space.\\

We decided to use a notion of overlap between distributions ($\BC$) to compare trajectories in the latent space, mainly for interpretation purposes. 
If willing to have a metric to compare distributions, many exist in closed-form for independent multivariate Gaussian. For instance, $\BC$ is related to the Hellinger distance by the following relationship: $H(p, q) = (1 - \BC(p, q))^{1/2}$.

We also set a threshold $s$ on the $\BC$s to define a high overlap in the latent space. In terms of comparing individual behaviours, we are just interested in proximity, as having two different or very different individual behaviour do not change the qualitative analysis (for instance if a trajectory is going far to the west, it does not matter if the other trajectory is going to the south or to the east, both are the fruits of different behaviours than the first trajectory). Hence, a need to binarize the indices. 
If willing to use network analysis tools to quantify the collective behaviour and group the fishing vessels without using binary networks,  would be to use a unimodal continuous distribution on $(0,1)$ (e.g. a Beta distribution with constrained parameters or a continuous Bernoulli distribution \citet{loaiza2019continuous}) as the emission distribution for the edges.

Finally, we set a quantile $q$ when building similarity networks over a defined time period.
A solution would be to study networks constructed with the same temporal scale as the data, i.e. daily proximity graphs.
Acknowledging the necessity to distinguish between noise or temporary local effects and collaborative effects, we plan on extending existing approaches to reconstruct networks from unreliable observations \citep{le2018estimating, young2020bayesian}, by accounting for the assortative community structure bias induced by the use of proximity graphs. 
It will help us to better account for collaborative effects in similarity networks, which can further enhance the reliability and accuracy of the analysis. The analysis would still need to be conducted at an appropriate time period to define a fishing fleet.

\section*{Aknowledgements}
This work was financially supported by France Filière Pêche (FFP) as part of the MACCOLOC project.

%\bibliographystyle{abbrvnat}
%\bibliography{references}

\begin{thebibliography}{39}
\providecommand{\natexlab}[1]{#1}
\providecommand{\url}[1]{\texttt{#1}}
\expandafter\ifx\csname urlstyle\endcsname\relax
  \providecommand{\doi}[1]{doi: #1}\else
  \providecommand{\doi}{doi: \begingroup \urlstyle{rm}\Url}\fi

\bibitem[Alglave et~al.(2022)Alglave, Rivot, Etienne, Woillez, Thorson, and
  Vermard]{alglave2022combining}
B.~Alglave, E.~Rivot, M.-P. Etienne, M.~Woillez, J.~T. Thorson, and Y.~Vermard.
\newblock Combining scientific survey and commercial catch data to map fish
  distribution.
\newblock \emph{ICES Journal of Marine Science}, 79\penalty0 (4):\penalty0
  1133--1149, 2022.
\newblock \doi{10.1093/icesjms/fsac032}.

\bibitem[Bengio et~al.(2013)Bengio, Courville, and
  Vincent]{bengio2013representation}
Y.~Bengio, A.~Courville, and P.~Vincent.
\newblock Representation learning: A review and new perspectives.
\newblock \emph{IEEE transactions on pattern analysis and machine
  intelligence}, 35\penalty0 (8):\penalty0 1798--1828, 2013.
\newblock \doi{10.1109/TPAMI.2013.50}.

\bibitem[Bez(2003)]{bez2003catchability}
N.~Bez.
\newblock The catchability in a spatial context: a simulation exercise.
\newblock In \emph{ICES CM 2003/X:09}, 2003.
\newblock URL
  \url{https://www.ices.dk/sites/pub/CM%20Doccuments/2003/X/X0903.PDF}.

\bibitem[Bhattacharyya(1943)]{bhattacharyya1943measure}
A.~Bhattacharyya.
\newblock On a measure of divergence between two statistical populations
  defined by their probability distribution.
\newblock \emph{Bulletin of the Calcutta Mathematical Society}, 35:\penalty0
  99--110, 1943.

\bibitem[Bishop(1998)]{bishop1998bayesian}
C.~M. Bishop.
\newblock Bayesian pca.
\newblock In \emph{Proceedings of the 11th International Conference on Neural
  Information Processing Systems}, NIPS'98, page 382–388, Cambridge, MA, USA,
  1998. MIT Press.

\bibitem[Blei et~al.(2017)Blei, Kucukelbir, and McAuliffe]{blei2017variational}
D.~M. Blei, A.~Kucukelbir, and J.~D. McAuliffe.
\newblock Variational inference: A review for statisticians.
\newblock \emph{Journal of the American statistical Association}, 112\penalty0
  (518):\penalty0 859--877, 2017.
\newblock \doi{10.1080/01621459.2017.1285773}.

\bibitem[Bénard et~al.(2022)Bénard, Veiga, and Scornet]{benard2022mean}
C.~Bénard, S.~D. Veiga, and E.~Scornet.
\newblock Mean decrease accuracy for random forests: inconsistency, and a
  practical solution via the sobol-mda.
\newblock \emph{Biometrika}, 109\penalty0 (4):\penalty0 881--900, 2022.
\newblock URL
  \url{https://EconPapers.repec.org/RePEc:oup:biomet:v:109:y:2022:i:4:p:881-900.}

\bibitem[Capobianco et~al.(2021)Capobianco, Millefiori, Forti, Braca, and
  Willett]{capobianco2021}
S.~Capobianco, L.~M. Millefiori, N.~Forti, P.~Braca, and P.~Willett.
\newblock Deep learning methods for vessel trajectory prediction based on
  recurrent neural networks.
\newblock \emph{IEEE Transactions on Aerospace and Electronic Systems},
  57\penalty0 (6):\penalty0 4329--4346, 12 2021.
\newblock \doi{10.1109/TAES.2021.3096873}.

\bibitem[Chabert-Liddell et~al.(In press)Chabert-Liddell, Barbillon, and
  Donnet]{chabertliddell2023}
S.-C. Chabert-Liddell, P.~Barbillon, and S.~Donnet.
\newblock Learning common structures in a collection of networks. an
  application to food webs.
\newblock \emph{The Annals of Applied Statistics}, In press.
\newblock URL \url{https://arxiv.org/abs/2206.00560}.

\bibitem[Fulton et~al.(2011)Fulton, Smith, Smith, and
  Van~Putten]{fulton2011human}
E.~A. Fulton, A.~D. Smith, D.~C. Smith, and I.~E. Van~Putten.
\newblock Human behaviour: the key source of uncertainty in fisheries
  management.
\newblock \emph{Fish and fisheries}, 12\penalty0 (1):\penalty0 2--17, 2011.
\newblock \doi{10.1111/j.1467-2979.2010.00371.x}.

\bibitem[Gao et~al.(2022)Gao, Xue, Shao, Zhao, Qin, Prabowo, Rahaman, and
  Salim]{gao2022generative}
N.~Gao, H.~Xue, W.~Shao, S.~Zhao, K.~K. Qin, A.~Prabowo, M.~S. Rahaman, and
  F.~D. Salim.
\newblock Generative {Adversarial} {Networks} for {Spatio}-temporal {Data}: {A}
  {Survey}.
\newblock \emph{ACM Transactions on Intelligent Systems and Technology},
  13\penalty0 (2):\penalty0 1--25, Apr. 2022.
\newblock ISSN 2157-6904, 2157-6912.
\newblock \doi{10.1145/3474838}.
\newblock URL \url{https://dl.acm.org/doi/10.1145/3474838}.

\bibitem[Graser et~al.(2023)Graser, Jalali, Lampert, {Weißenfeld}, and
  Janowicz]{graser2023}
A.~Graser, A.~Jalali, J.~Lampert, A.~{Weißenfeld}, and K.~Janowicz.
\newblock Deep learning from trajectory data: a review of deep neural networks
  and the trajectory data representations to train them.
\newblock In \emph{Workshop on Big Mobility Data Analysis BMDA2023 in
  conjunction with EDBT/ICDT}, 2023.

\bibitem[Higgins et~al.(2017)Higgins, Matthey, Pal, Burgess, Glorot, Botvinick,
  Mohamed, and Lerchner]{higgins2017betavae}
I.~Higgins, L.~Matthey, A.~Pal, C.~Burgess, X.~Glorot, M.~Botvinick,
  S.~Mohamed, and A.~Lerchner.
\newblock beta-{VAE}: Learning basic visual concepts with a constrained
  variational framework.
\newblock In \emph{International Conference on Learning Representations}, 2017.
\newblock URL \url{https://openreview.net/forum?id=Sy2fzU9gl}.

\bibitem[Hinz et~al.(2013)Hinz, Murray, Lambert, Hiddink, and Kaiser]{hinz2013}
H.~Hinz, L.~G. Murray, G.~I. Lambert, J.~G. Hiddink, and M.~J. Kaiser.
\newblock Confidentiality over fishing effort data threatens science and
  management progress.
\newblock \emph{Fish and Fisheries}, 14\penalty0 (1):\penalty0 110--117, Mar.
  2013.
\newblock ISSN 1467-2979.
\newblock \doi{10.1111/j.1467-2979.2012.00475.x}.
\newblock URL \url{http://dx.doi.org/10.1111/j.1467-2979.2012.00475.x}.

\bibitem[Holland et~al.(1983)Holland, Laskey, and Leinhardt]{holland1983}
P.~W. Holland, K.~B. Laskey, and S.~Leinhardt.
\newblock Stochastic blockmodels: First steps.
\newblock \emph{Social Networks}, 5\penalty0 (2):\penalty0 109--137, 06 1983.
\newblock \doi{10.1016/0378-8733(83)90021-7}.
\newblock URL
  \url{https://linkinghub.elsevier.com/retrieve/pii/0378873383900217}.

\bibitem[ICES(2022)]{ices2022working}
ICES.
\newblock Working group for the bay of biscay and the iberian waters ecoregion
  (wgbie), 10 2022.
\newblock URL
  \url{https://ices-library.figshare.com/articles/report/Working_Group_for_the_Bay_of_Biscay_and_the_Iberian_Waters_Ecoregion_WGBIE_/20068988}.

\bibitem[Jia et~al.(2022)Jia, Ma, He, Su, Zhang, and Yu]{jia2022motion}
C.~Jia, J.~Ma, M.~He, Y.~Su, Y.~Zhang, and Q.~Yu.
\newblock Motion primitives learning of ship-ship interaction patterns in
  encounter situations.
\newblock \emph{Ocean Engineering}, 247:\penalty0 110708, 2022.
\newblock \doi{10.1016/j.oceaneng.2022.110708}.

\bibitem[Joo et~al.(2018)Joo, Etienne, Bez, and {Mahévas}]{joo2018}
R.~Joo, M.-P. Etienne, N.~Bez, and S.~{Mahévas}.
\newblock Metrics for describing dyadic movement: a review.
\newblock \emph{Movement Ecology}, 6\penalty0 (1):\penalty0 26, 12 2018.
\newblock \doi{10.1186/s40462-018-0144-2}.
\newblock URL \url{https://doi.org/10.1186/s40462-018-0144-2}.

\bibitem[Joo et~al.(2021)Joo, Bez, Etienne, Marin, Goascoz, Roux, and
  {Mahévas}]{joo2021}
R.~Joo, N.~Bez, M.-P. Etienne, P.~Marin, N.~Goascoz, J.~Roux, and
  S.~{Mahévas}.
\newblock Identifying partners at sea from joint movement metrics of pelagic
  pair trawlers.
\newblock \emph{ICES Journal of Marine Science}, 78\penalty0 (5):\penalty0
  1758--1768, 09 2021.
\newblock \doi{10.1093/icesjms/fsab068}.
\newblock URL \url{https://academic.oup.com/icesjms/article/78/5/1758/6276504}.

\bibitem[Kingma and Ba(2015)]{kingma2015adam}
D.~P. Kingma and J.~Ba.
\newblock Adam: {A} method for stochastic optimization.
\newblock In Y.~Bengio and Y.~LeCun, editors, \emph{3rd International
  Conference on Learning Representations, {ICLR} 2015, San Diego, CA, USA, May
  7-9, 2015, Conference Track Proceedings}, 2015.
\newblock URL \url{http://arxiv.org/abs/1412.6980}.

\bibitem[Kingma and Welling(2014)]{kingma2014}
D.~P. Kingma and M.~Welling.
\newblock Auto-encoding variational bayes.
\newblock In Y.~Bengio and Y.~LeCun, editors, \emph{2nd International
  Conference on Learning Representations, {ICLR} 2014, Banff, AB, Canada, April
  14-16, 2014, Conference Track Proceedings}, 2014.
\newblock URL \url{http://arxiv.org/abs/1312.6114}.

\bibitem[Kramer(1991)]{kramer1991nonlinear}
M.~A. Kramer.
\newblock Nonlinear principal component analysis using autoassociative neural
  networks.
\newblock \emph{AIChE journal}, 37\penalty0 (2):\penalty0 233--243, 1991.

\bibitem[Lawrence and Hyv{\"a}rinen(2005)]{lawrence2005probabilistic}
N.~Lawrence and A.~Hyv{\"a}rinen.
\newblock Probabilistic non-linear principal component analysis with gaussian
  process latent variable models.
\newblock \emph{Journal of machine learning research}, 6\penalty0 (11), 2005.
\newblock URL \url{http://jmlr.org/papers/v6/lawrence05a.html}.

\bibitem[Le et~al.(2018)Le, Levin, and Levina]{le2018estimating}
C.~M. Le, K.~Levin, and E.~Levina.
\newblock {Estimating a network from multiple noisy realizations}.
\newblock \emph{Electronic Journal of Statistics}, 12\penalty0 (2):\penalty0
  4697 -- 4740, 2018.
\newblock \doi{10.1214/18-EJS1521}.
\newblock URL \url{https://doi.org/10.1214/18-EJS1521}.

\bibitem[Lee and Wilkinson(2019)]{lee2019review}
C.~Lee and D.~J. Wilkinson.
\newblock A review of stochastic block models and extensions for graph
  clustering.
\newblock \emph{Applied Network Science}, 4\penalty0 (1):\penalty0 1--50, 2019.
\newblock \doi{10.1007/s41109-019-0232-2}.

\bibitem[Liang et~al.(2021)Liang, Liu, Li, Xiao, Liu, and
  Lu]{liang2021unsupervised}
M.~Liang, R.~W. Liu, S.~Li, Z.~Xiao, X.~Liu, and F.~Lu.
\newblock An unsupervised learning method with convolutional auto-encoder for
  vessel trajectory similarity computation.
\newblock \emph{Ocean Engineering}, 225:\penalty0 108803, 2021.
\newblock \doi{10.1016/j.oceaneng.2021.108803}.

\bibitem[Loaiza-Ganem and Cunningham(2019)]{loaiza2019continuous}
G.~Loaiza-Ganem and J.~P. Cunningham.
\newblock The continuous bernoulli: fixing a pervasive error in variational
  autoencoders.
\newblock \emph{Advances in Neural Information Processing Systems}, 32, 2019.

\bibitem[Olive et~al.(2020)Olive, Basora, Viry, and Alligier]{olive2020deep}
X.~Olive, L.~Basora, B.~Viry, and R.~Alligier.
\newblock Deep trajectory clustering with autoencoders.
\newblock In \emph{ICRAT 2020, 9th International Conference for Research in Air
  Transportation}, 2020.

\bibitem[Paszke et~al.(2019)Paszke, Gross, Massa, Lerer, Bradbury, Chanan,
  Killeen, Lin, Gimelshein, Antiga, et~al.]{pytorch}
A.~Paszke, S.~Gross, F.~Massa, A.~Lerer, J.~Bradbury, G.~Chanan, T.~Killeen,
  Z.~Lin, N.~Gimelshein, L.~Antiga, et~al.
\newblock Pytorch: An imperative style, high-performance deep learning library.
\newblock \emph{Advances in neural information processing systems}, 32, 2019.

\bibitem[Poos and Rijnsdorp(2007)]{poos2007experiment}
J.-J. Poos and A.~D. Rijnsdorp.
\newblock An "experiment" on effort allocation of fishing vessels: the role of
  interference competition and area specialization.
\newblock \emph{Canadian Journal of Fisheries and Aquatic Sciences},
  64\penalty0 (2):\penalty0 304--313, 2007.
\newblock \doi{10.1139/f06-177}.

\bibitem[Radford et~al.(2016)Radford, Metz, and Chintala]{radford2016}
A.~Radford, L.~Metz, and S.~Chintala.
\newblock Unsupervised representation learning with deep convolutional
  generative adversarial networks.
\newblock In Y.~Bengio and Y.~LeCun, editors, \emph{4th International
  Conference on Learning Representations, {ICLR} 2016, San Juan, Puerto Rico,
  May 2-4, 2016, Conference Track Proceedings}, 2016.
\newblock URL \url{http://arxiv.org/abs/1511.06434}.

\bibitem[Rezende et~al.(2014)Rezende, Mohamed, and
  Wierstra]{rezende2014stochastic}
D.~J. Rezende, S.~Mohamed, and D.~Wierstra.
\newblock Stochastic backpropagation and approximate inference in deep
  generative models.
\newblock In \emph{International conference on machine learning}, pages
  1278--1286. PMLR, 2014.

\bibitem[Roux et~al.(2023)Roux, Bez, Rochet, Joo, and {Mahévas}]{roux2023}
J.~Roux, N.~Bez, P.~Rochet, R.~Joo, and S.~{Mahévas}.
\newblock Graphlet correlation distance to compare small graphs.
\newblock \emph{PLOS ONE}, 18\penalty0 (2):\penalty0 e0281646, 02 2023.
\newblock \doi{10.1371/journal.pone.0281646}.
\newblock URL
  \url{https://journals.plos.org/plosone/article?id=10.1371/journal.pone.0281646}.
\newblock Publisher: Public Library of Science.

\bibitem[Roy et~al.(2022)Roy, Fablet, and Bertrand]{roy2022}
A.~Roy, R.~Fablet, and S.~L. Bertrand.
\newblock Using generative adversarial networks (gan) to simulate central-place
  foraging trajectories.
\newblock \emph{Methods in Ecology and Evolution}, 13\penalty0 (6):\penalty0
  1275--1287, 2022.
\newblock \doi{10.1111/2041-210X.13853}.
\newblock URL
  \url{https://onlinelibrary.wiley.com/doi/abs/10.1111/2041-210X.13853}.

\bibitem[Skafte et~al.(2019)Skafte, J{\o}rgensen, and
  Hauberg]{skafte2019reliable}
N.~Skafte, M.~J{\o}rgensen, and S.~Hauberg.
\newblock Reliable training and estimation of variance networks.
\newblock \emph{Advances in Neural Information Processing Systems}, 32, 2019.

\bibitem[Svensson et~al.(2020)Svensson, Gayoso, Yosef, and
  Pachter]{svensson2020interpretable}
V.~Svensson, A.~Gayoso, N.~Yosef, and L.~Pachter.
\newblock Interpretable factor models of single-cell rna-seq via variational
  autoencoders.
\newblock \emph{Bioinformatics}, 36\penalty0 (11):\penalty0 3418--3421, 2020.
\newblock \doi{10.1093/bioinformatics/btaa169}.

\bibitem[Van~der Maaten and Hinton(2008)]{van2008visualizing}
L.~Van~der Maaten and G.~Hinton.
\newblock Visualizing data using t-sne.
\newblock \emph{Journal of machine learning research}, 9\penalty0 (11), 2008.

\bibitem[Wright and Ziegler(2017)]{ranger}
M.~N. Wright and A.~Ziegler.
\newblock {ranger}: A fast implementation of random forests for high
  dimensional data in {C++} and {R}.
\newblock \emph{Journal of Statistical Software}, 77\penalty0 (1):\penalty0
  1--17, 2017.
\newblock \doi{10.18637/jss.v077.i01}.

\bibitem[Young et~al.(2020)Young, Cantwell, and Newman]{young2020bayesian}
J.-G. Young, G.~T. Cantwell, and M.~Newman.
\newblock Bayesian inference of network structure from unreliable data.
\newblock \emph{Journal of Complex Networks}, 8\penalty0 (6):\penalty0 cnaa046,
  2020.
\newblock \doi{10.1093/comnet/cnaa046}.

\end{thebibliography}

\appendix

\section{Convolutional neural network architecture}\label{app:sec:cnn}

We provide the architecture used in this paper for the decoder and the encoder with covariates for $d_\latent = 3$ as Pytorch pseudocode:

%\begin{multicols}{2}

\begin{verbatim}
    Encoder:
        Convolutional layer:
            Conv1d(2+2, 8, kernel_size=4, stride=2)
            Batchnorm1d(8)
            LeakyReLU(0.2)
        Convolutional layer:
            Conv1d(8, 32, kernel_size=4, stride=2)
            Batchnorm1d(32)
            LeakyReLU(0.2)
        Convolutional layer:
            Conv1d(32, 128, kernel_size=3, stride=2)
            Batchnorm1d(128)
            LeakyReLU(0.2)
        Convolutional layer:
            Conv1d(128, 3, kernel_size=2, stride=1)

    Decoder:
        Convolutional layer:
            Conv1d(3+2, 20, kernel_size=2, stride=1)
            Batchnorm1d(20)
            LeakyReLU(0.2)
        Strided convolutional layer:
            ConvTranspose1d(20, 10, kernel_size=5, stride=2)
            Batchnorm1d(10)
            ReLU()
        Strided convolutional layer:
            ConvTranspose1d(10, 5, kernel_size=4, stride=2)
            Batchnorm1d(5)
            ReLU()
        Strided convolutional layer:
            ConvTranpose1d(5, 2, kernel_size=2, stride=1)
\end{verbatim}

\section{Details of the probabilistic model for proximity graphs}\label{app:sec:sbm}

Let $B_{(q,y)}$ be the number of vessels having done at least a fishing trip on quarter $(q,y)$. The vessels are divided into $R$ groups, which drive their connections. Let $(W^{(q,y)}_{b}) = r$ if vessel $b$ belongs to group $r \in \{1,\dots, R\}$ in quarter $(q,y)$. The group memberships are independently distributed with probability:
\begin{equation*}
    p\left(W^{(q,y)}_{b}=r\right) = \rho_{r},
\end{equation*}
where $(\rho_r)_{r \in \llbracket1;R\rrbracket})$ is a probability vector.
Given their respective group memberships, the connections between each couple of vessels $(b,b')$, $b < b'$ are assumed to be independent and distributed as:
\begin{equation*}
   A^{(q,y)}_{bb'} | W^{(q,y)}_{br}W^{(q,y)}_{b'r'} = 1 \sim \mathcal{B}(\alpha_{rr'}),
\end{equation*}
where $\mathcal{B}$ is the Bernoulli distribution. As the networks are undirected, we set $A^{(q,y)}_{b'b}=A^{(q,y)}_{bb'}$.
The above defines the SBM model, $SBM_{B_{(q,y)}}(R, \alpha, \rho)$. Then the collection of networks is distributed as:
\begin{equation*}
    \mathbf{A} = \prod_{(q,y)}A^{(q,y)}, \quad \mbox{where} \quad A^{(q,y)} \sim \mbox{SBM}_{B_{(q,y)}} (R, \alpha, \rho).
\end{equation*}

For $R$ fixed, the inference of the model parameters and the latent group are done through the maximization of a variational approximation of the log-likelihood. A penalized model selection criterion is used to choose the number of groups $\widehat{R}$ \citep{chabertliddell2023}.

\section{Quarterly proximity graphs}\label{app:sec:proxy-graph}

\begin{figure}[!h]
    \centering
    \includegraphics[width=\textwidth]{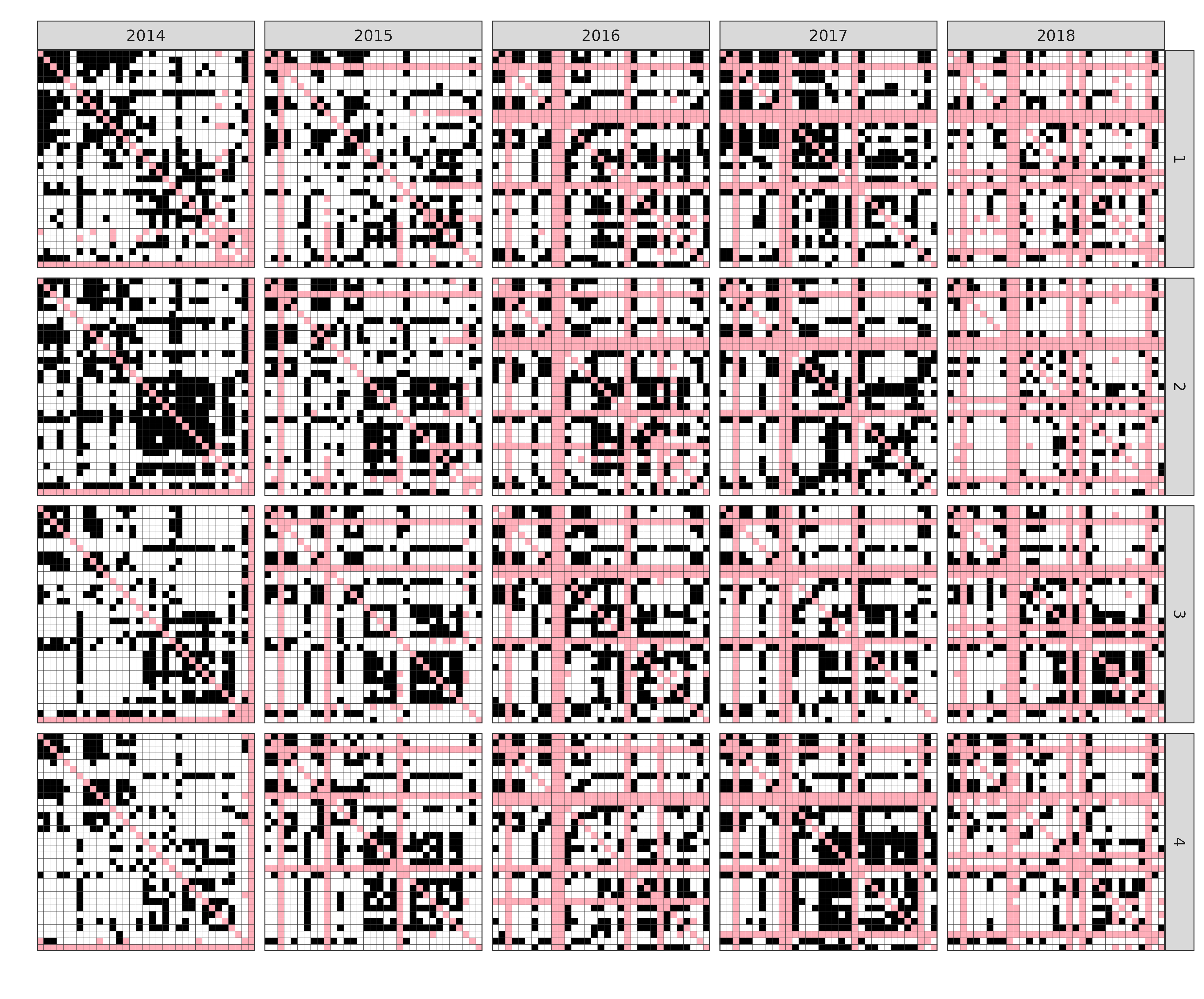}
    \caption{Adjacency matrices of the $20$ quarterly proximity graphs between fishing vessels. Matrices are ordered by quarter (row) and year (column). Proximity between vessels is printed in black and \texttt{NA} value in red.}
    \label{fig:proxy-graph-trim}
\end{figure}

\end{document}